# Consistent determination of geometrically necessary dislocation density from simulations and experiments

Suchandrima Das[a], Felix Hofmann[a], Edmund Tarleton[b]*

[a]*Department of Engineering Science, University of Oxford, Parks Road, Oxford OX1 3PJ, UK*

[b]*Department of Materials, University of Oxford, Parks Road, Oxford OX1 3PH, UK*

*correspondence should be addressed to edmund.tarleton@materials.ox.ac.uk

# Abstract

The use of Nye's dislocation tensor for calculating the density of geometrically necessary dislocations (GND) is widely adopted in the study of plastically deformed materials. The "curl" operation involved in finding the Nye tensor, while conceptually straightforward has been marred with inconsistencies and several different definitions are in use. For the three most common definitions, we show that their consistent application leads to the same result. To eliminate frequently encountered confusion, a summary of expressions for Nye's tensor in terms of elastic and plastic deformation gradient, and for both small and large deformations, is presented. A further question when estimating GND density concerns the optimization technique used to solve the under-determined set of equations linking Nye's tensor and GND density. A systematic comparison of the densities obtained by two widely used techniques, $L^1$ and $L^2$ minimisation, shows that both methods yield remarkably similar total GND densities. Thus the mathematically simpler, $L^2$, may be preferred over $L^1$ except when information about the distribution of densities on specific slip systems is required. To illustrate this, we compare experimentally measured lattice distortions beneath nano-indents in pure tungsten, probed using 3D-resolved synchrotron X-ray micro-diffraction, with those predicted by 3D strain-gradient crystal plasticity finite element calculations. The results are in good agreement and show that the volumetric component of the elastic strain field has a surprisingly small effect on the determined Nye tensor. This is important for experimental techniques, such as micro-beam Laue measurements and HR-EBSD, where only the deviatoric strain component is measured.

# 1. Introduction

Before the development of crystal plasticity finite element formulations, phenomenological continuum models were used to describe plastic deformation in materials. Without delving into the underlying microstructural processes, these approaches, calibrated by experiments, could capture plasticity at the macroscopic scale (Dunne and Petrinic, 2004; Khan, Akhtar.S, 1995). Their great advantage is simplicity, however a lack of physical basis severely limits their predictive capabilities, especially for processes where microstructural heterogeneity is important.

Crystal plasticity finite element (CPFE) formulations address this issue by explicitly modelling plasticity in terms of crystallographic slip at the grain scale (Roters et al., 2010). Popularity of these formulations has increased dramatically as they directly account for complex interactions between individual grains of polycrystals and the resulting locally heterogeneous loading. Beginning from 1982, when it was first introduced by Peirce et al. (Peirce et al., 1982), the CPFE technique has developed to span a range of constitutive and numerical formulations, applicable to a large number of problems. For example CPFE has been used to simulate the development of microstructures and the consequent effect on the macroscopic material response (Aifantis, 1984), to simulate surface roughening in thin film mechanics problems (Raabe et al., 2003), grain-boundary and interface mechanics (Bate and Hutchinson, 2005; Meissonnier et al., 2001), strain-gradient effects (Dunne et al., 2012, 2007), polycrystalline morphology and texture, the necessary conditions (energy) for crack nucleation (Chen et al., 2017), geometrically necessary dislocations (GNDs) (Dahlberg et al., 2014), creep and high temperature deformation (Balasubramanian and Anand, 2002), texture formation (Asaro and Rice, 1977), deformation twinning (Kalidindi, 1998), multiphase mechanics (Vogler and Clayton, 2008) etcetera. Importantly CPFE methods can be easily adapted to different material systems, simply by modifying the crystallographic slip law. CPFE has been applied to a diverse range of not only metals (Balasubramanian and Anand, 2002; Dunne et al., 2012; Li et al., 2009; Vogler and Clayton, 2008), but also rocks (Behrmann, 1985). Furthermore CPFE has been used across a range of length scales, from single crystals (Wang et al., 2004; Weber et al., 2008) to polycrystals (Zhao et al., 2008) and multiscale applications combined with ab initio calculations (Raabe et al., 2007). CPFE has also been used with other modelling techniques such as continuum dislocation dynamics, and nonlinear thermoelasticity to simulate the response of materials under extreme dynamic loading (Luscher et al., 2016).

In the early 70s, empirical viscoplastic formulations were primarily used, based on the "flow-potential" approach proposed by Rice (Rice, 1971) for time-dependent plastic deformation. Subsequent works of Rice and Asaro (Asaro and Rice, 1977) and Peirce and Needleman (Peirce et al., 1982), focused on the analysis of non-uniform, localized deformation of ductile crystals, where crystal slip was simulated by a rate-independent, elastic-plastic relation, following the Schmid law. Owing to computational restrictions, their simulations involved a simplified scenario of a single slip, or two symmetric slip systems.

With increasing computational power, a wide range of microstructure-based multi-scale plasticity models has emerged. These include various grain- and sub-grain scale problems, as well as complex 2D, 3D grain morphologies. The introduction of strain gradient terms in the constitutive law marked a major step forward, making it possible to capture experimentally observed size effects. Numerous strain gradient plasticity (SGP) formulations have been proposed, for example by Fleck and Hutchinson (Fleck and Hutchinson, 1997; Fleck et al., 1994), Gao, Huang, Nix and Hutchinson (Gao et al., 1999), Arsenlis and Parks (Arsenlis and Parks, 1999), Cheong and Busso (Busso et al., 2000), Gurtin and Anand (Gurtin and Anand, 2005a, 2005b), Dunne et al. (Dunne et al., 2007) and Fleck and Willis (N A Fleck and Willis, 2009; N. A. Fleck and Willis, 2009). These approaches have enabled accurate simulations of inelastic, scale-dependent deformation phenomena such as the increasing hardness of metals and ceramics with decreasing indenter size in indentation simulations (Wang et al., 2004), or Hall-Petch grain size strengthening effect (Lim et al., 2014; Lyu et al., 2015; Raabe et al., 2003). SGP phenomenological formulations have also been extended to calculate the fraction of the rate of plastic work converted into heat, by taking into account a strain dependent factor to include the locked in strain energy around statistically stored dislocations (Lubarda, 2016).

In particular, SGP formulations helped to numerically simulate the length-scale effects and production of GNDs associated with non-uniform plastic deformation. Smaller characteristic length-scales lead to steeper strain gradients and hence higher GND densities, causing a size effect as flow stress depends on dislocation density (Nye, 1953). Liu et al. (Liu et al., 2013) compared experimental and theoretical evaluations (using tension and torsion on polycrystalline copper wires) of three phenomenological theories of strain gradient plasticity, to show that the size effects seen in plastic flow is primarily due to the GND density generated as a result of plastic strain gradients. Using both mechanism-based and phenomenological SGP theories, Paneda et al. (Martínez-Pañeda and Niordson, 2016) showed localized strain hardening near crack tips, promoted by GNDs. Compelling as the results are, experimental techniques are required to confirm these numerical simulations. The critical thickness theory has recently been used to get a more reasonable estimate of the length-scale (in the µm range) from the underlying fundamental physical quantities to facilitate the use of the SGP theory in engineering applications such as finite element applications (Liu and Dunstan, 2017).

In strain gradient CPFE formulations, plastic deformation is accounted for by dislocation glide on active slip systems and the deformation gradient is linked to the lattice curvature and in turn to the additional GNDs, generated in the slip systems, to accommodate this lattice curvature. The length-scale effect within the concept of GNDs is captured here by using Nye's formulation of the dislocation tensor (Nye, 1953). In a recent study, Lyu et al. (Lyu et al., 2015) modelled crystal plasticity using a continuum dislocation dynamics model (CDD) and used this together with a viscoplastic self-consistent model to study the evolution of dislocation densities in multi-phase steels.

To calculate GND density, the closure failure of a suitable Burgers' circuit can be considered (Ashby, 1970; Nye, 1953). Using Stokes theorem, this can be recast as the computation of the "curl" of the deformation gradient to form the dislocation or Nye tensor (Nye, 1953). Whilst straightforward conceptually, this step has been marred with inconsistencies. A review of the literature reporting GND density calculations shows a wide range of different curl definitions being used, often with erroneous applications of ± signs, misplaced indices and missing transpose operations. These errors will lead to unphysical results. For example a missing transpose operation effectively corresponds to swapping of Burgers' vector and line direction, resulting in incorrect dislocation densities. If a minus sign is erroneously placed then left handed screw densities become right handed and edge dislocation densities have their extra half plane on the opposite side of the slip plane. The first key goal of this paper is to compare the three most commonly used curl definitions and to establish the correct expressions to be used.

A further question in the computation of GND density concerns the optimization technique used to solve the under-determined set of equations linking the curl of the deformation gradient and GND density. Two optimization techniques, $L^1$ and $L^2$, are commonly employed. Each yields a different solution and very few studies (Wallis et al., 2017; Wilkinson and Randman, 2010) have investigated the differences in the results they produce. Here we carry out a systematic comparison of the GND densities predicted by both methods to determine their applicability in different scenarios.

For the validation of strain gradient CPFE models, a direct comparison to experiments performed at the same length-scale is essential. Here we present experimental measurements and strain gradient CPFE calculations of the lattice distortions beneath a spherical nano-indent in a tungsten single crystal. Experimental techniques such as electron back-scattered diffraction (EBSD), high-resolution EBSD (HR-EBSD), high-resolution digital image correlation (HR-DIC) are commonly used to measure lattice distortions in two-dimensions, for example studies by (Kysar et al., 2010), (Dahlberg et al., 2014), (Kartal et al., 2015), (Zhang et al., 2016), (Ruggles et al., 2016), (Guan et al., 2017) etcetera. Barabash et al. (Barabash et al., 2009) showed how GNDs and the effective strain gradient change the white beam Laue patterns of shocked materials. With the aim of capturing the GND formation in plastic deformation, we use the synchrotron X-ray micro-Laue diffraction technique to non-destructively probe the full lattice rotation and residual elastic strain field with 3D spatial resolution, and without altering the residual stress state. Using strain gradient CPFE calculations we carry out a detailed 3D simulation of the same experiment. The experimentally measured and predicted lattice rotations, strains and GND densities are compared in detail. This gives rise to several interesting questions, for example concerning the effect of the volumetric elastic strain on GND density calculations, since Laue diffraction only measures the deviatoric lattice strain tensor (Chung and Ice, 1999). This question is examined by modelling the experimentally recorded data using the strain gradient CPFE calculations.

We begin by describing the different definitions for calculating the curl of a second-order tensor. This is followed by a review of the theoretical framework of the computation of the dislocation tensor and GND density. Based on this the expressions for the dislocation tensor in terms of elastic or plastic deformation gradient, as well as lattice strains and rotations, are discussed. Next, a comparison of nano-indentation-induced lattice distortions measured by Laue diffraction and predicted by strain gradient CPFE simulations is presented. Finally, the effects of $L^1$ or $L^2$ optimisation techniques, and volumetric elastic strain on the computation of GND densities are explored.

# 2. Theoretical Background

## 2.1 Computing the Curl of a second-order tensor

As noted by Robert W. Soutas (Soutas-Little, 1999), there are several different definitions in use for computing the curl of a second order tensor. Here, three different approaches to the curl computation are discussed. Importantly we show that, if used consistently, they all lead to the same end result.

Let $\boldsymbol{P}$ and $\boldsymbol{V}$ be general second-order tensors. The $km$ component of the pre-curl of $\boldsymbol{P}$ is denoted as $(\nabla \times \boldsymbol{P})_{km}$, while the post-curl is $(\boldsymbol{V} \times \nabla)_{km}$ (Soutas-Little, 1999). In component form these may be stated as

$$\text{Precurl: } R_{km} = (\nabla \times \boldsymbol{P})_{km} = \epsilon_{ijk} P_{jm,i} \tag{1}$$

$$\text{Postcurl: } S_{km} = (\boldsymbol{V} \times \nabla)_{km} = -\epsilon_{ijm} V_{kj,i} \tag{2}$$

where, $\epsilon_{ijk}$ is the permutation tensor. An alternative third definition (referred to as "curl3" throughout this text) is commonly used in computational crystal plasticity studies, e.g. by Arsenlis and Parks (Arsenlis and Parks, 1999) and Cermelli and Gurtin (Cermelli and Gurtin, 2001):

$$\text{curl3: } Q_{km} = (\nabla \times \boldsymbol{V})_{km} = \epsilon_{ijk} V_{mj,i} \tag{3}$$

The derivation of each of these three curl formulae, in tensor notation, is shown in Appendix A. **Comparing the precurl (Eq. (1)) and curl3 (Eq. (3)), it can be seen that $R_{km}$, the precurl of $\boldsymbol{P}$, will be equal $Q_{km}$, the curl3 of $\boldsymbol{V}$, when $\boldsymbol{V} = \boldsymbol{P}^{\mathrm{T}}$.**

$$R_{km} = (\nabla \times \boldsymbol{P})_{km} = \epsilon_{ijk} P_{jm,i} = \epsilon_{ijk} P^{\mathrm{T}}_{mj,i} = \epsilon_{ijk} V_{mj,i} = (\nabla \times \boldsymbol{V})_{km} = Q_{km} \tag{4}$$

**The post-curl definition is the negative transpose of curl3**. This can be shown as

$$\text{Postcurl: } S_{km} = (\boldsymbol{V} \times \nabla)_{km} = -\epsilon_{ijm} V_{kj,i} = -(\epsilon_{ijk} V_{mj,i})^{T} = -(Q_{km})^{T}. \tag{5}$$

Explicit versions of Eq. (4) and (5), in component form, are provided in Appendix B. In summary, for any second-order tensor $\boldsymbol{V}$, the curl computation using each of the three discussed conventions, may be equated as

$$\mathbf{curl3}\ (\boldsymbol{V}) = \mathbf{Precurl}\left(\boldsymbol{V}^{\mathrm{T}}\right) = -(\mathbf{Postcurl}\ (\boldsymbol{V}))^{\mathrm{T}} \tag{6}$$

## 2.2 Calculation of dislocation tensor

A geometrical link between the lattice curvature and the distribution of GNDs is given by the dislocation tensor (also known as the Nye tensor), $\boldsymbol{\alpha}^{Nye}$, proposed by Nye (Nye, 1953). Nye initially formulated $\boldsymbol{\alpha}^{Nye}$ using only the lattice rotation gradients, assuming that no long-range elastic strain fields are present. Kroner and Ashby further developed this approach, by adding elastic strain gradients to the formulation of the dislocation tensor (Arsenlis and Parks, 1999). Below a summary of this theory is provided.

The analysis below, for small deformations, closely follows the derivation by Fleck and Hutchinson (Fleck et al., 1994) in their study establishing the concept of strain gradient plasticity. Figure 1b shows a representative crystal lattice within an imaginary solid, with a chosen Cartesian reference frame as depicted. We assume that plastic flow occurs by dislocation motion and that the lattice is stretched and rotated during elastic deformation.

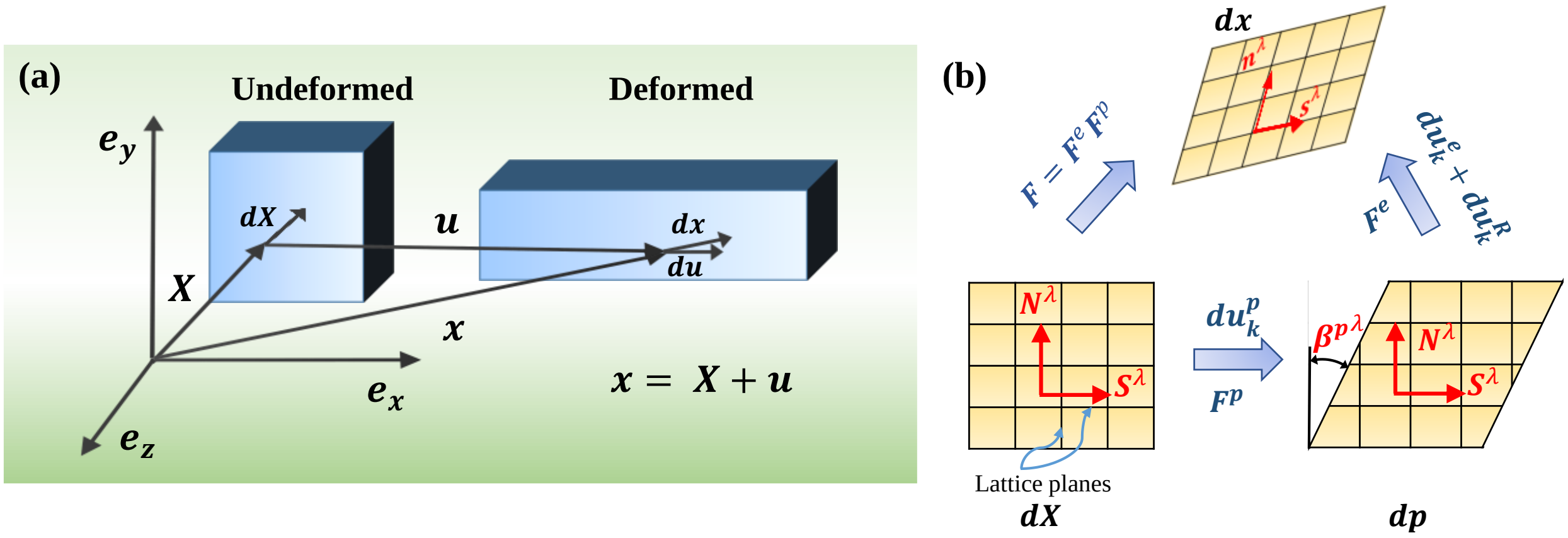

Figure 1- (a) Schematic representation of the deformation of a body. $\boldsymbol{u}$ contains the information about whether the deformation is a translation, rigid body rotation or stretch, or a combination of these.
(b) Schematic showing the multiplicative decomposition of the deformation gradient as the crystal lattice deforms from the initial state $dX$ to the intermediate imaginary state $dp$ (where only plastic deformation has taken place) and finally to state $dx$.

Let us consider the relative displacement, $\boldsymbol{du}$, of two material points which are separated by $\boldsymbol{dX}$ as shown in Figure 1a. This relative displacement can be split into three components (Fleck et al., 1994) as follows

$$\boldsymbol{du} = \boldsymbol{du}^p + \boldsymbol{du}^R + \boldsymbol{du}^e \tag{7}$$

where,

$$du_k^p = \beta_{kj}^p dX_j \tag{8}$$

$$du_k^R = \omega_{kj} dX_j \tag{9}$$

$$du_k^e = \varepsilon_{kj}^{el} dX_j \tag{10}$$

$\boldsymbol{du}^p$ is the relative displacement caused by slip and described by the slip tensor $\boldsymbol{\beta}^p$ . $\boldsymbol{du}^R$ is caused by lattice rotation $\boldsymbol{\omega}$ and $\boldsymbol{du}^e$ is due to elastic strain, $\boldsymbol{\varepsilon}^{el}$. For a specific slip systems, $\lambda$, defined by slip direction $\boldsymbol{s}^\lambda$ and slip plane normal $\boldsymbol{n}^\lambda$, and crystallographic slip, $\beta^{p\lambda}$, the slip tensor, $\boldsymbol{\beta}^p$ , is given by the contribution from each of the active slip systems. Thus,

$$\boldsymbol{\beta}^p = \sum_\lambda \beta^{p\lambda} \boldsymbol{s}^\lambda \otimes \boldsymbol{n}^\lambda \tag{11}$$

Following Nye's reasoning (Nye, 1953) the closure failure of a Burgers' circuit, *c*, on surface *S* with plane normal $\boldsymbol{N}$ (see Figure E.1 (b)) can be used to link the crystallographic slip to the resultant Burgers' vector, $< \boldsymbol{B} >$:

$$< B >_k = \oint_c du_k^p = \oint_c \beta_{kj}^p dX_j \tag{12}$$

Using Stokes' theorem this can be rewritten as:

$$< B >_k = \iint_S (\in_{ijm} \beta_{kj,i}^p).N_m \, dS = \iint_S \alpha_{km}^{Nye}.N_m \, dS \tag{13}$$

where $\boldsymbol{\alpha}^{Nye}$ is the dislocation tensor defined by Nye. Thus, the Nye tensor corresponds to the curl of the slip tensor (Fleck et al., 1994):

$$\alpha_{km}^{Nye} = \in_{ijm} \beta_{kj,i}^{p} \tag{14}$$

Nye (Nye, 1953) related $\boldsymbol{\alpha}^{Nye}$ to the dislocation distribution inside the crystal. Given $q$ dislocations per unit area with Burgers' vector $\boldsymbol{b}$ and unit line direction $\boldsymbol{l}$ threading the plane, with surface unit normal $\boldsymbol{N}$, $\boldsymbol{\alpha}^{Nye}$ may be written as

$$\alpha_{km}^{Nye} = q\, b_k l_m \tag{15}$$

Defining $\rho_m = q\, l_m$ means $\alpha_{km}^{Nye} = b_k \rho_m$. So, Nye's dislocation tensor may be written as

$$\boldsymbol{\alpha}^{Nye} = \sum_{\lambda} (\boldsymbol{b}^{\lambda} \otimes \boldsymbol{\rho}^{\lambda}) \tag{16}$$

where $\lambda$ is a general slip system.

Here, it is important to note that the total displacement, $d\boldsymbol{u}$, along any closed circuit must be zero. Thus, the closure failure brought about by crystallographic slip, $< B >_k = \oint_c du_k^p$, has to be balanced by an equal and opposite displacement incompatibility, i.e. $\oint_c (du_k^R + du_k^e)$. This relation between the closure failure due to slip (i.e. plastic displacement) and that due to elastic displacement can be described by the concept of the deformation gradient as outline below:

Figure 1a shows an imaginary material in the undeformed configuration with a line vector $\boldsymbol{dX}$. After deformation, this line vector is transformed to $\boldsymbol{dx}$. From here on we distinguish between undeformed and deformed coordinate systems. **Variables in upper case correspond to undeformed coordinates, while lower case refers to the deformed coordinates.**

The deformation gradient, $\boldsymbol{F}$, can be defined as a second-order tensor that maps the undeformed state to the deformed state of a sample. This can be written as

$$\boldsymbol{F} = \frac{\partial \boldsymbol{x}}{\partial \boldsymbol{X}} = \boldsymbol{I} + \frac{\partial \boldsymbol{u}}{\partial \boldsymbol{X}} = \boldsymbol{I} + \boldsymbol{\beta} \tag{17}$$

where, $\boldsymbol{u}$ is the displacement, $\boldsymbol{I}$ is the identity matrix and $\boldsymbol{\beta}$ the displacement gradient. $\boldsymbol{\beta}$ contains information about whether the deformation involves a rigid body rotation or a shape change or both (Dunne and Petrinic, 2004). When the deformation includes both elastic and plastic contributions, the deformation gradient can be split into elastic ($\boldsymbol{F}^e$) and plastic ($\boldsymbol{F}^p$) parts, where each of the gradients may contain both stretch and rigid body rotation.

In Eq. (14) $\boldsymbol{\alpha}^{Nye}$, is defined as a curl computation of the slip tensor. The plastic deformation gradient $\boldsymbol{F}^p$ captures the deformation by crystallographic slip, which is the same as the deformation captured by the slip tensor (Eq. (8)). In fact, $\boldsymbol{F}^p = \boldsymbol{I} + \boldsymbol{\beta}^p$. Hence Nye's dislocation tensor may be written in terms of $\boldsymbol{F}^p$ as $\alpha_{mk}^{Nye} = \in_{ijm} F_{kj,i}^p$.

Kroner (Kroner, 1955) and Bilby (Lazar and Pellegrini, 2016) expressed the dislocation density tensor as the negative of the expression adopted by Nye (Eq. 14). Consequently, Arsenlis and Parks (Arsenlis and Parks, 1999) and Cermelli and Gurtin (Cermelli and Gurtin, 2001) used the curl3 convention to find the dislocation density tensor from the plastic deformation gradient. Rewriting $\boldsymbol{\alpha}^{Nye}$ (Eq. (14)) in terms of the curl3 formula gives

$$\alpha_{km}^{Nye} = \in_{ijm} F_{kj,i}^p = \left(\in_{ijk} F_{mj,i}^p\right)^T = ((\mathrm{curl3}\ (\boldsymbol{F}^p))^T)_{km} \tag{18}$$

**From here onwards, unless otherwise specified, the curl operation signifies performing the curl computation using the curl3 convention**. Owing to the contribution made by several researchers to the concept of the dislocation density tensor, from here on we adopt the notation of $\boldsymbol{\alpha}$ to represent it. Eq. (18) can also be arrived at considering the following approach: To separate the elastic ($\boldsymbol{F}^e$) and the plastic ($\boldsymbol{F}^p$) deformations, an intermediate imaginary configuration $d\boldsymbol{p}$ can be introduced (Figure 1b), where the sample has undergone purely plastic deformation. The transformation of $d\boldsymbol{X}$ to $d\boldsymbol{p}$ is captured by the plastic deformation gradient ($\boldsymbol{F}^p$)

$$d\boldsymbol{p} = \boldsymbol{F}^p d\boldsymbol{X} \tag{19}$$

$d\boldsymbol{p}$ can then be mapped to the vector $d\boldsymbol{x}$ by the elastic deformation gradient

$$d\boldsymbol{x} = \boldsymbol{F}^e d\boldsymbol{p} \tag{20}$$

and equating Eq. (19) and (20)

$$d\boldsymbol{p} = \boldsymbol{F}^{p} d\boldsymbol{X} = \boldsymbol{F}^{e^{-1}} d\boldsymbol{x} \tag{21}$$

Rewriting Eq. (21) the multiplicative decomposition of the deformation gradient is obtained as proposed by Lee et al. (Lee, 1969)

$$\boldsymbol{F} = \frac{\partial \boldsymbol{x}}{\partial \boldsymbol{X}} = \boldsymbol{F}^{e} \boldsymbol{F}^{p}. \tag{22}$$

Substituting $\boldsymbol{F}^{e}$ and $\boldsymbol{F}^{p}$ into Eq. (17) and introducing elastic and plastic parts of the displacement gradient, $\boldsymbol{\beta}^{e}$ and $\boldsymbol{\beta}^{p}$ respectively, gives

$$\boldsymbol{F} = I + \boldsymbol{\beta} = \boldsymbol{F}^{e} \boldsymbol{F}^{p} = (I + \boldsymbol{\beta}^{e})(I + \boldsymbol{\beta}^{p}) = I + \boldsymbol{\beta}^{e} + \boldsymbol{\beta}^{p} + \boldsymbol{\beta}^{e} \boldsymbol{\beta}^{p}. \tag{23}$$

For small deformations, the $\boldsymbol{\beta}^{e} \boldsymbol{\beta}^{p}$ term is negligible and thus the displacement gradient for small strains may be written as

$$\boldsymbol{\beta} \cong \boldsymbol{\beta}^{e} + \boldsymbol{\beta}^{p}. \tag{24}$$

From the kinematics of deformation, closure failure of a region can be defined as the change in length of a path on the surface due to generation of dislocations in the volume. Acharya and Bassani (Acharya and Bassani, 2000) defined closure failure with respect to deformed configuration $\boldsymbol{x}$, where $\boldsymbol{x} = \boldsymbol{X} + \boldsymbol{u}$, as

$$< \boldsymbol{b} >= \oint_{c} \boldsymbol{F}^{e^{-1}} d\boldsymbol{x} \tag{25}$$

where <$\boldsymbol{b}$> is the net Burgers vector of the dislocation lines passing through closed loop $c$. This is analogous to Eq. (12) in the deformed configuration. Using Stokes theorem, the integration around $c$ may be replaced by integration over any surface patch $s$, bounded by $c$ and with plane normal $\boldsymbol{n}$, so that

$$< \boldsymbol{b} >= \oint_c \boldsymbol{F}^{e-1} d\boldsymbol{x} = \iint_s (\text{curl}\, \boldsymbol{F}^{e-1})^T \boldsymbol{n} ds \cong \iint_s (-\text{curl}\, \boldsymbol{F}^{e})^T \boldsymbol{n} ds \tag{26}$$

where, curl $\boldsymbol{F}^{e-1} \cong -\text{curl}\, \boldsymbol{F}^{e}$ for small elastic strains because

$$\boldsymbol{F}^{e-1} = (\boldsymbol{I} + \boldsymbol{\beta}^{e})^{-1} \cong \boldsymbol{I} - \boldsymbol{\beta}^{e} \text{ so curl}\, \boldsymbol{F}^{e-1} \cong -\text{curl}\, \boldsymbol{F}^{e} \tag{27}$$

as curl of the identity matrix is zero. Re-writing in terms of the displacement gradient,

$$< \boldsymbol{b} >= \iint_s (\text{curl}\, (\boldsymbol{F}^{e-1}))^T \boldsymbol{n} ds \cong \iint_s (-\text{curl}\, (\boldsymbol{\beta}^{e}))^T \boldsymbol{n} ds \tag{28}$$

The transpose in Eq. (28) is introduced when applying Stokes' theorem to higher order tensors, as proved by Cermelli and Gurtin (details in Appendix C) (Cermelli and Gurtin, 2001). The closure failure, represented in terms of the non-vanishing cumulative Burgers' vector of all dislocations, can also be written in terms of the undeformed configuration $\boldsymbol{X}$. Computation of curl in the undeformed configuration, $d\boldsymbol{X}$, will be referred to as "CURL" from here on.

$$< \boldsymbol{b} >= \oint_c \boldsymbol{F}^{e-1} d\boldsymbol{x} = \oint_c \boldsymbol{F}^{e-1} \boldsymbol{F} d\boldsymbol{X} = \oint_c \boldsymbol{F}^{e-1} \boldsymbol{F}^{e} \boldsymbol{F}^{p} d\boldsymbol{X} = \oint_C \boldsymbol{F}^{p} d\boldsymbol{X} = \iint_S (\text{CURL}\, \boldsymbol{F}^{p})^T \boldsymbol{N} dS = < \boldsymbol{B} > \tag{29}$$

Here <***b***> and <***B***> refers to the resultant Burgers' vector in the deformed and undeformed coordinate frame respectively. Rewriting Eq. (29) in terms of the plastic displacement gradient,

$$< \boldsymbol{B} >= \iint_S (\text{CURL}\, (I + \boldsymbol{\beta}^{p}))^T \boldsymbol{N} dS = \iint_S (\text{CURL}\, (\boldsymbol{\beta}^{p}))^T \boldsymbol{N} dS \tag{30}$$

Thus, in summary, the closure failure can be represented in terms of either the elastic or plastic displacement gradient as

$$< \boldsymbol{B} >= \iint_S (\text{CURL}\, (\boldsymbol{\beta}^{p}))^T \boldsymbol{N} dS \cong \iint_s \left(-\text{CURL}\, (\boldsymbol{\beta}^{e})\right)^T \boldsymbol{N} dS \cong \iint_s \left(-\text{curl}\, (\boldsymbol{\beta}^{e})\right)^T \boldsymbol{n} ds = < \boldsymbol{b} > \tag{31}$$

where for small deformation we do not need to distinguish between the initial and deformed system. Equating Eq. (16) and Eq. (31), the dislocation tensor may be re-written to again arrive at Eq. (18).

$$\boldsymbol{\alpha} = \sum_i (\boldsymbol{b}^i \otimes \boldsymbol{\rho}^i) = (\mathrm{CURL}\,(\,\boldsymbol{F}^p))^T \;= (\mathrm{CURL}\,(\,\boldsymbol{\beta}^p))^T \cong (-\mathrm{curl}\,(\boldsymbol{\beta}^e))^T \cong (-\mathrm{CURL}\,(\boldsymbol{\beta}^e))^T \tag{32}$$

The elastic strain, $\boldsymbol{\varepsilon}^e$, and lattice rotation, $\boldsymbol{\omega}^e$, are related to the displacement tensor by

$$\boldsymbol{\varepsilon}^e = \frac{1}{2}\,(\boldsymbol{\beta}^e + \boldsymbol{\beta}^{e^T});\ \boldsymbol{\omega}^e = \frac{1}{2}\,(\boldsymbol{\beta}^e - \boldsymbol{\beta}^{e^T}) \tag{33}$$

Thus, Eq. (32) may further be re-written as

$$\boldsymbol{\alpha} \cong (-\mathrm{curl}(\,\boldsymbol{\varepsilon}^e + \,\boldsymbol{\omega}^e))^T \cong (-\mathrm{CURL}(\,\boldsymbol{\varepsilon}^e + \,\boldsymbol{\omega}^e))^T \tag{34}$$

for the case of small deformation.

Given the deformation gradients, $\boldsymbol{F}^e$ or $\boldsymbol{F}^p$, or displacement gradients, $\boldsymbol{\beta}^p$ or $\boldsymbol{\beta}^e$, $\boldsymbol{\alpha}$ can be computed using any of the three curl definitions discussed above (Eq. (1), (2) & (3)). Following Eq. (6), a summary of expressions for $\boldsymbol{\alpha}$ in terms of the three curl definitions is provided in Table 1.

| $\boldsymbol{\alpha}$ | In terms of:<br><br>CURL ($\boldsymbol{\beta}^p$) or CURL ($\boldsymbol{F}^p$) for all deformations | In terms of:<br><br>curl ($\boldsymbol{\beta}^e$) for small deformations<br><br>curl ($\boldsymbol{F}^{e^{-1}}$) for large deformations |
|---|---|---|
| Using<br><br>"curl3" | $(\text{CURL}\,(\boldsymbol{F}^p))^T = (\text{CURL}\,(\boldsymbol{\beta}^p))^T$<br><br>where, $\text{CURL}\,(\boldsymbol{\beta}^p)_{km} = \in_{ijk}\,\beta^p_{mj,i}$ | $(\text{curl}\,(\boldsymbol{F}^{e^{-1}}))^T \cong (-\text{curl}\,(\boldsymbol{\beta}^e))^T$<br><br>where, $\text{curl}\,(\boldsymbol{\beta}^e)_{km} = \in_{ijk}\,\beta^e_{mj,i}$<br><br>$\text{curl}\,(\boldsymbol{F}^{e^{-1}})_{km} = \in_{ijk}\,F^{e^{-1}}_{mj,i}$ |
| Using<br><br>"Pre-curl" | $\left((\text{CURL}\,(\boldsymbol{F}^{p\mathrm{T}})\right)^T = (\text{CURL}\,(\boldsymbol{\beta}^{p^T}))^T$<br><br>where, $\text{CURL}(\boldsymbol{\beta}^{p^T})_{km} = \in_{ijk}\,(\boldsymbol{\beta}^{p^T})_{jm,i}$ | $(\text{curl}\,\boldsymbol{F}^{e-1^T})^{\mathrm{T}} \cong (-\text{curl}\,(\boldsymbol{\beta}^{e^T}))^T$<br><br>where,$\text{curl}\,(\boldsymbol{\beta}^{e^T})_{km} = \in_{ijk}\,(\boldsymbol{\beta}^{e^T})_{jm,i}$<br><br>$\text{curl}\,(\boldsymbol{F}^{e-1^T})_{km} = \in_{ijk}\,(\boldsymbol{F}^{e-1^T})_{jm,i}$ |
| Using<br><br>"Post-curl" | $(-(\text{CURL}\,(\boldsymbol{F}^p))^T)^T = -\text{CURL}\,(\boldsymbol{F}^p)$<br>$= -\text{CURL}\,(\boldsymbol{\beta}^p)$<br><br>where, $\text{CURL}\,(\boldsymbol{\beta}^p)_{km} = -\in_{ijm}\,\beta^p_{kj,i}$ | $\left(-\left(\text{curl}\,(\boldsymbol{F}^{e^{-1}})\right)^T\right)^T = -\text{curl}\,(\boldsymbol{F}^{e^{-1}})$<br>$\cong \text{curl}\,(\boldsymbol{\beta}^e)$<br><br>where, $\text{curl}\,(\boldsymbol{\beta}^e)_{km} = -\in_{ijm}\,\beta^e_{kj,i}$<br><br>$\text{curl}\,(\boldsymbol{F}^{e^{-1}}) = \in_{ijm}\,F^{e^{-1}}_{kj,i}$ |

Table 1 – Summary of computation of dislocation tensor using the three common different definitions of curl in terms of $\boldsymbol{\beta}$ and $\boldsymbol{F}$.

For small strains the dislocation tensor may be explicitly written in terms of the plastic ($\boldsymbol{\beta}^p$) or elastic ($\boldsymbol{\beta}^e$) displacement gradients:

$$\boldsymbol{\alpha} \cong \begin{bmatrix} \frac{\partial \beta_{13}^p}{\partial x_2} - \frac{\partial \beta_{12}^p}{\partial x_3} & \frac{\partial \beta_{11}^p}{\partial x_3} - \frac{\partial \beta_{13}^p}{\partial x_1} & \frac{\partial \beta_{12}^p}{\partial x_1} - \frac{\partial \beta_{11}^p}{\partial x_2} \\ \frac{\partial \beta_{23}^p}{\partial x_2} - \frac{\partial \beta_{22}^p}{\partial x_3} & \frac{\partial \beta_{21}^p}{\partial x_3} - \frac{\partial \beta_{23}^p}{\partial x_1} & \frac{\partial \beta_{22}^p}{\partial x_1} - \frac{\partial \beta_{21}^p}{\partial x_2} \\ \frac{\partial \beta_{33}^p}{\partial x_2} - \frac{\partial \beta_{32}^p}{\partial x_3} & \frac{\partial \beta_{31}^p}{\partial x_3} - \frac{\partial \beta_{33}^p}{\partial x_1} & \frac{\partial \beta_{32}^p}{\partial x_1} - \frac{\partial \beta_{31}^p}{\partial x_2} \end{bmatrix} \tag{35}$$

$$\boldsymbol{\alpha} \cong \begin{bmatrix} \frac{\partial \beta_{12}^e}{\partial x_3} - \frac{\partial \beta_{13}^e}{\partial x_2} & \frac{\partial \beta_{13}^e}{\partial x_1} - \frac{\partial \beta_{11}^e}{\partial x_3} & \frac{\partial \beta_{11}^e}{\partial x_2} - \frac{\partial \beta_{12}^e}{\partial x_1} \\ \frac{\partial \beta_{22}^e}{\partial x_3} - \frac{\partial \beta_{23}^e}{\partial x_2} & \frac{\partial \beta_{23}^e}{\partial x_1} - \frac{\partial \beta_{21}^e}{\partial x_3} & \frac{\partial \beta_{21}^e}{\partial x_2} - \frac{\partial \beta_{22}^e}{\partial x_1} \\ \frac{\partial \beta_{32}^e}{\partial x_3} - \frac{\partial \beta_{33}^e}{\partial x_2} & \frac{\partial \beta_{33}^e}{\partial x_1} - \frac{\partial \beta_{31}^e}{\partial x_3} & \frac{\partial \beta_{31}^e}{\partial x_2} - \frac{\partial \beta_{32}^e}{\partial x_1} \end{bmatrix} \tag{36}$$

The displacement gradient can also be written in terms of the lattice rotation and lattice strain as

$$\boldsymbol{\beta}^e = \boldsymbol{\omega}^e + \boldsymbol{\varepsilon}^e \tag{37}$$

This leads to the following form for the dislocation tensor:

$$\boldsymbol{\alpha} \cong \begin{bmatrix} \frac{\partial \omega_{12}^e}{\partial x_3} - \frac{\partial \omega_{13}^e}{\partial x_2} & \frac{\partial \omega_{13}^e}{\partial x_1} & \frac{\partial \omega_{21}^e}{\partial x_1} \\ \frac{\partial \omega_{32}^e}{\partial x_2} & \frac{\partial \omega_{23}^e}{\partial x_1} - \frac{\partial \omega_{21}^e}{\partial x_3} & \frac{\partial \omega_{21}^e}{\partial x_2} \\ \frac{\partial \omega_{32}^e}{\partial x_3} & \frac{\partial \omega_{13}^e}{\partial x_3} & \frac{\partial \omega_{31}^e}{\partial x_2} - \frac{\partial \omega_{32}^e}{\partial x_1} \end{bmatrix} + \begin{bmatrix} \frac{\partial \varepsilon_{12}^e}{\partial x_3} - \frac{\partial \varepsilon_{13}^e}{\partial x_2} & \frac{\partial \varepsilon_{13}^e}{\partial x_1} - \frac{\partial \varepsilon_{11}^e}{\partial x_3} & \frac{\partial \varepsilon_{11}^e}{\partial x_2} - \frac{\partial \varepsilon_{12}^e}{\partial x_1} \\ \frac{\partial \varepsilon_{22}^e}{\partial x_3} - \frac{\partial \varepsilon_{23}^e}{\partial x_2} & \frac{\partial \varepsilon_{23}^e}{\partial x_1} - \frac{\partial \varepsilon_{21}^e}{\partial x_3} & \frac{\partial \varepsilon_{21}^e}{\partial x_2} - \frac{\partial \varepsilon_{22}^e}{\partial x_1} \\ \frac{\partial \varepsilon_{32}^e}{\partial x_3} - \frac{\partial \varepsilon_{33}^e}{\partial x_2} & \frac{\partial \varepsilon_{33}^e}{\partial x_1} - \frac{\partial \varepsilon_{31}^e}{\partial x_3} & \frac{\partial \varepsilon_{31}^e}{\partial x_2} - \frac{\partial \varepsilon_{32}^e}{\partial x_1} \end{bmatrix} \tag{38}$$

where, considering the asymmetric nature of $\boldsymbol{\omega}^e$, the $\omega_{ii}^e$ components are 0. In general, the lattice rotation gradients are substantially greater than the elastic strain gradients and make a more significant contribution to the dislocation tensor.

## 2.3 Calculation of GND density

Eq. (16) relates the dislocation tensor to the densities of geometrically necessary dislocations. The 3×3 $\boldsymbol{\alpha}$ tensor can be reshaped as a 9×1 column vector. A linear operator $\boldsymbol{A}$ is formed (9×$j$ matrix, for $j$ types of dislocations), where the $j^{th}$ column contains the dyadic product of the Burgers' vector and line direction of the $j^{th}$ dislocation type. Representing the densities of the $j$ dislocation types as a column vector $\boldsymbol{\rho}$, the quantities $\boldsymbol{\alpha}$, $\boldsymbol{\rho}$ and $\boldsymbol{A}$ may be used to recast Eq. (16) as

$$\boldsymbol{A\rho} = \boldsymbol{\alpha} \tag{39}$$

Explicitly this may be written as

$$\sum_j (\boldsymbol{b}^j \otimes \boldsymbol{\rho}^j) = \begin{bmatrix} \boldsymbol{b}_1^1\boldsymbol{l}_1^1 & \boldsymbol{b}_1^2\boldsymbol{l}_1^2 & \boldsymbol{b}_1^3\boldsymbol{l}_1^3 & . & . & . & . & . & . & . & \boldsymbol{b}_1^j\boldsymbol{l}_1^j \\ \boldsymbol{b}_1^1\boldsymbol{l}_2^1 & \boldsymbol{b}_1^2\boldsymbol{l}_2^2 & \boldsymbol{b}_1^3\boldsymbol{l}_2^3 & . & . & . & . & . & . & . & \boldsymbol{b}_1^j\boldsymbol{l}_2^j \\ \boldsymbol{b}_1^1\boldsymbol{l}_3^1 & \boldsymbol{b}_1^2\boldsymbol{l}_3^2 & \boldsymbol{b}_1^3\boldsymbol{l}_3^3 & . & . & . & . & . & . & . & \boldsymbol{b}_1^j\boldsymbol{l}_3^j \\ \boldsymbol{b}_2^1\boldsymbol{l}_1^1 & \boldsymbol{b}_2^2\boldsymbol{l}_1^2 & \boldsymbol{b}_2^3\boldsymbol{l}_1^3 & . & . & . & . & . & . & . & \boldsymbol{b}_2^j\boldsymbol{l}_1^j \\ \boldsymbol{b}_2^1\boldsymbol{l}_2^1 & \boldsymbol{b}_2^2\boldsymbol{l}_2^2 & \boldsymbol{b}_2^3\boldsymbol{l}_2^3 & . & . & . & . & . & . & . & \boldsymbol{b}_2^j\boldsymbol{l}_2^j \\ \boldsymbol{b}_2^1\boldsymbol{l}_3^1 & \boldsymbol{b}_2^2\boldsymbol{l}_3^2 & \boldsymbol{b}_2^3\boldsymbol{l}_3^3 & . & . & . & . & . & . & . & \boldsymbol{b}_2^j\boldsymbol{l}_3^j \\ \boldsymbol{b}_3^1\boldsymbol{l}_1^1 & \boldsymbol{b}_3^2\boldsymbol{l}_1^2 & \boldsymbol{b}_3^3\boldsymbol{l}_1^3 & . & . & . & . & . & . & . & \boldsymbol{b}_3^j\boldsymbol{l}_1^j \\ \boldsymbol{b}_3^1\boldsymbol{l}_2^1 & \boldsymbol{b}_3^2\boldsymbol{l}_2^2 & \boldsymbol{b}_3^3\boldsymbol{l}_2^3 & . & . & . & . & . & . & . & \boldsymbol{b}_3^j\boldsymbol{l}_2^j \\ \boldsymbol{b}_3^1\boldsymbol{l}_3^1 & \boldsymbol{b}_3^2\boldsymbol{l}_3^2 & \boldsymbol{b}_3^3\boldsymbol{l}_3^3 & . & . & . & . & . & . & . & \boldsymbol{b}_3^j\boldsymbol{l}_3^j \end{bmatrix} \begin{bmatrix} \boldsymbol{\rho}_1 \\ \boldsymbol{\rho}_2 \\ \boldsymbol{\rho}_3 \\ . \\ . \\ . \\ . \\ . \\ . \\ . \\ . \\ . \\ . \\ . \\ . \\ \boldsymbol{\rho}_j \end{bmatrix} = \begin{bmatrix} \boldsymbol{\alpha}_{11} \\ \boldsymbol{\alpha}_{12} \\ \boldsymbol{\alpha}_{13} \\ \boldsymbol{\alpha}_{21} \\ \boldsymbol{\alpha}_{22} \\ \boldsymbol{\alpha}_{23} \\ \boldsymbol{\alpha}_{31} \\ \boldsymbol{\alpha}_{32} \\ \boldsymbol{\alpha}_{33} \end{bmatrix} \tag{40}$$

where, $\boldsymbol{l}$ is the line direction. Since generally j>9 there is no unique solution for $\boldsymbol{\rho}$. Instead, knowing $\boldsymbol{\alpha}$ and $\boldsymbol{A}$, optimization methods may be used to obtain $\boldsymbol{\rho}$. The mathematically simplest is the $L^2$ optimization scheme (Arsenlis and Parks, 1999), which minimizes the sum of squares of dislocation densities i.e $\sum_j \rho_j^2 = \boldsymbol{\rho}^T.\boldsymbol{\rho}$. Using the right pseudo inverse, the solution may be written as

$$\boldsymbol{\rho} = \boldsymbol{A}^T(\boldsymbol{A}\boldsymbol{A}^T)^{-1}\boldsymbol{\alpha} \tag{41}$$

When using the $L^2$ optimization, it is essential to construct $\boldsymbol{A}$ including all possible slip systems not just the active ones (i.e. the calculation is independent of the resolved shear stress). For example, in a BCC crystal, if {110}<111> slip systems are being considered, all 16 possible dislocation types (12 edge + 4 screw) must be included, leading to an $\boldsymbol{A}$ matrix with dimensions [9x16]. The disadvantage of this optimisation scheme is a lack of any physical basis.

An alternative optimization method, here referred to as $L^1$, minimizes the total dislocation elastic energy, i.e $(1-\upsilon)^{-1}\sum_j \rho_j^{edge} + \sum_j \rho_j^{screw}$, to obtain a solution for Eq. (39). In the example above, since all the dislocations have the same Burgers vector magnitude, and assuming elastic isotropy, differences arise only due to the ratio of energies for edge and screw dislocations (Wilkinson and Randman, 2010):

$$\frac{E_{edge}}{E_{screw}} = \frac{1}{1-\nu} \tag{42}$$

where, ν is the Poisson's ratio. The "linprog" algorithm, implemented in Matlab (The Mathworks Inc.; www.mathworks.com) was used to perform the $L^1$ optimisation.

A key assumption is that dislocations are either pure edge or pure screw. Other methods for solving Eq. 40 involve minimising the total dislocation density (Demir et al., 2009; El-Dasher et al., 2003; Sun et al., 2000), or minimisation the equivalent line length (Wilkinson and Randman, 2010). A thorough, in-depth comparison all these minimisation norms is beyond the scope of this paper. Instead we focus on a comparison of the two most commonly used methods, $L^1$ and $L^2$ minimisations. For both these optimisations, the solution obtained represents only one of the infinite number of solutions to Eq. (39). The total dislocation density is obtained by summing the magnitudes of the densities of all *j* dislocation types.

# 3. Material and Methods

To illustrate the concepts discussed above, we now consider the lattice distortions, dislocation tensor and GND density beneath a spherical nano-indent in a pure tungsten single crystal. A direct comparison is made between experimental measurements and numerical predictions from a strain-gradient CPFE model of the indentation process.

A [001]-oriented high purity tungsten single crystal (99.99 wt.%) was mechanically polished using diamond paste and colloidal silica to produce a near defect-free mirror finish. 500 nm deep indents were made using a MTS NanoXp indenter with a spherical, ~4.2 μm radius diamond tip. Synchrotron

X-ray micro-beam Laue diffraction was used to probe the residual lattice distortions beneath a specific indent with sub-micron (~0.5 microns) 3D resolution. Briefly, micro-beam Laue diffraction measurements were carried out at beamline 34-ID-E, Advanced Photon Source, Argonne National Lab, USA. A polychromatic X-ray beam (7-30 keV) was focused by KB mirrors to a probe spot of ~500 nm full width at half maximum at the sample. The sample was placed at this probe spot in 45° reflection geometry and the orientation of the laboratory coordinates in relation to the initial crystallographic coordinates is shown by the X, Y, Z axes and their respective directions, superimposed on the sample image (Figure 2). Laue diffraction patterns were recorded by an area detector (Perkin-Elmer, #XRD 1621, with pixel size 200 × 200 µm) placed ~511 mm above the sample. Depth resolved measurements were made possible by using the differential aperture X-ray microscopy (DAXM) technique. Further details about the DAXM technique and the experimental data processing are provided in Appendix D and elsewhere (Das et al., 2018).

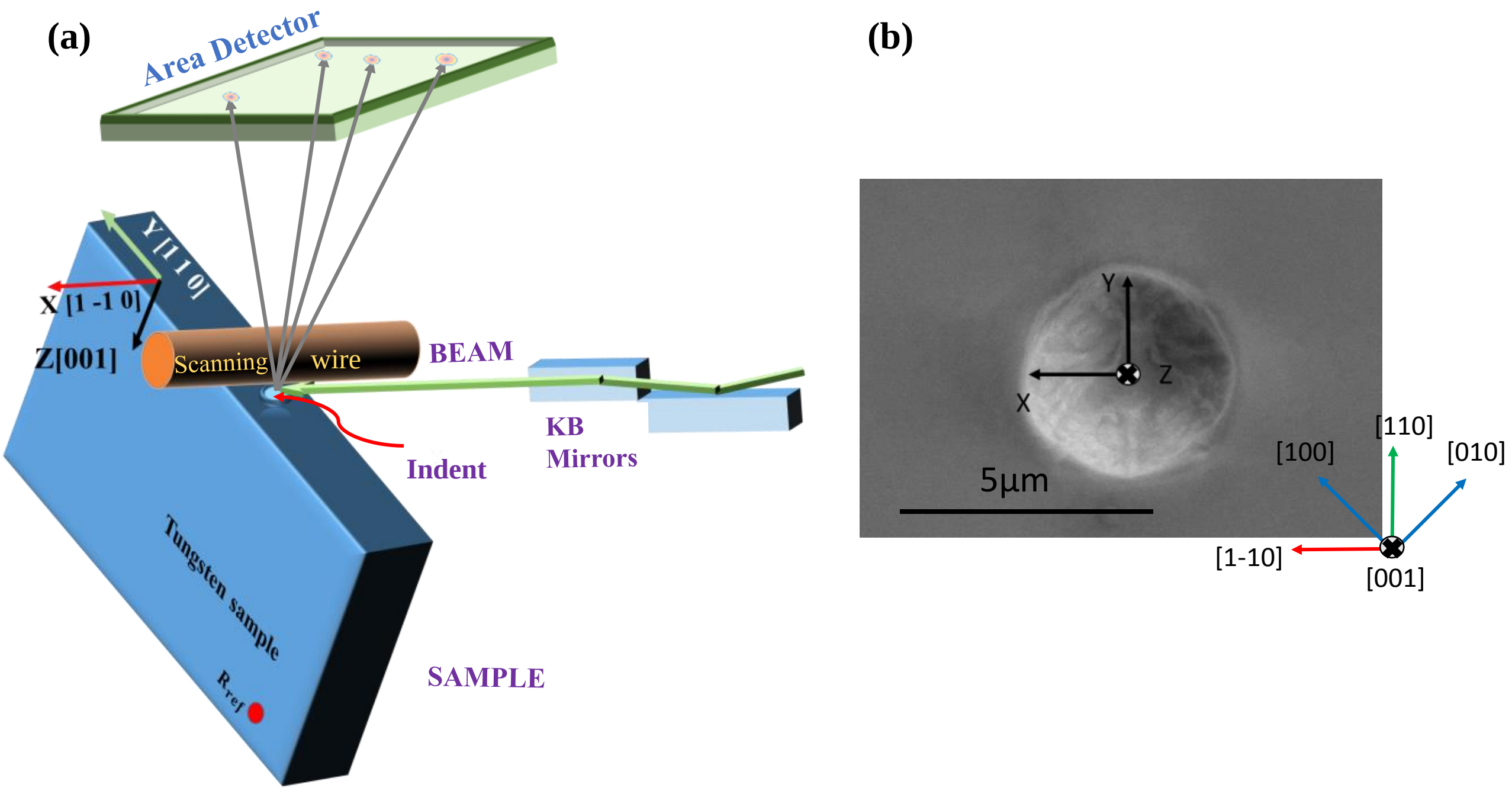

Figure 2 – (a) Schematic of the experimental Laue diffraction setup at beamline 34-ID-E at the APS. The sample is positioned at 45° reflection geometry and the orientation of the sample coordinates (X, Y, Z), in relation to the initial crystal axes is shown. (b) SEM image of the indent on the tungsten sample surface with the sample coordinate system superimposed.

A 3D finite element model was constructed to simulate the nano-indentation experiment in Abaqus (Dassault Systèmes, Providence, RI, USA). The indentation model (Figure 3) comprised of a 3D single crystal tungsten cube ($20 \times 20 \times 20\ \mu m^3$) representing one quarter of the experimental setup with elastic properties as stated in Table 2. Crystal plasticity was implemented in the model using a UMAT subroutine (details of the UMAT and the crystallographic slip law used is provided in Appendix E) and assumptions of isotropic elasticity and small deformations were made in the numerical simulation.

| $E_{tungsten}$ | $\nu_{tungsten}$ | $E_{diamond}$ | $\nu_{diamond}$ | $E_{eff}$ |
|---|---|---|---|---|
| 410 GPa | 0.28 | 1143 GPa | 0.0691 | 322.58 GPa |

Table 2 - Values of Young's modulus and Poisson's ratio for diamond (indenter tip) and tungsten (indented sample) as obtained from literature[1] (Ayres et al., 1975; Bolef and De Klerk, 1962; Featherston and Neighbours, 1963; Klein and Cardinale, 1993).

The boundary conditions imposed on the tungsten block included symmetric boundary conditions on the XZ and YZ surfaces near the indent, a traction free top surface, and fixed displacement and rotation boundary conditions on the remaining surfaces. The modelled spherical indenter (4.2 µm radius) was assumed to be a discrete rigid part, and contact between the tungsten block and the indenter was defined using the Abaqus node to surface contact algorithm. Consistent with the nano-indentation experiment, in the simulation, a displacement of 0.5 µm was applied to the indenter. A refined finite element mesh (applied edge bias 0.1 to 2 µm) with >15700 20-noded, reduced integration (8 integration point) 3D quadratic elements was used (C3D20R). The experimentally measured nano-indentation load-displacement data was used to refine the critically resolved shear stress parameter (CRSS), used in the slip law, to ensure accurate reproduction of the load-displacement curve. The effective modulus, $E_{eff}$, (Table E.2) was taken into consideration (Li et al.,

[1] With the assumption of an isotropic, linear elastic solid, the Young's modulus and Poisson's ratio are related to the elastic constant as follows: $E = c_{11} - 2\left(\frac{c_{12}^2}{c_{11} + c_{12}}\right)$ and $\nu = c_{12}/(c_{11} + c_{12})$.

2009) to account for the modulus of diamond and $\tau_c$ was rescaled accordingly (900 MPa). Figure 3 shows the Von Mises stress (after unloading) in the simulated model, mirrored about the YZ plane.

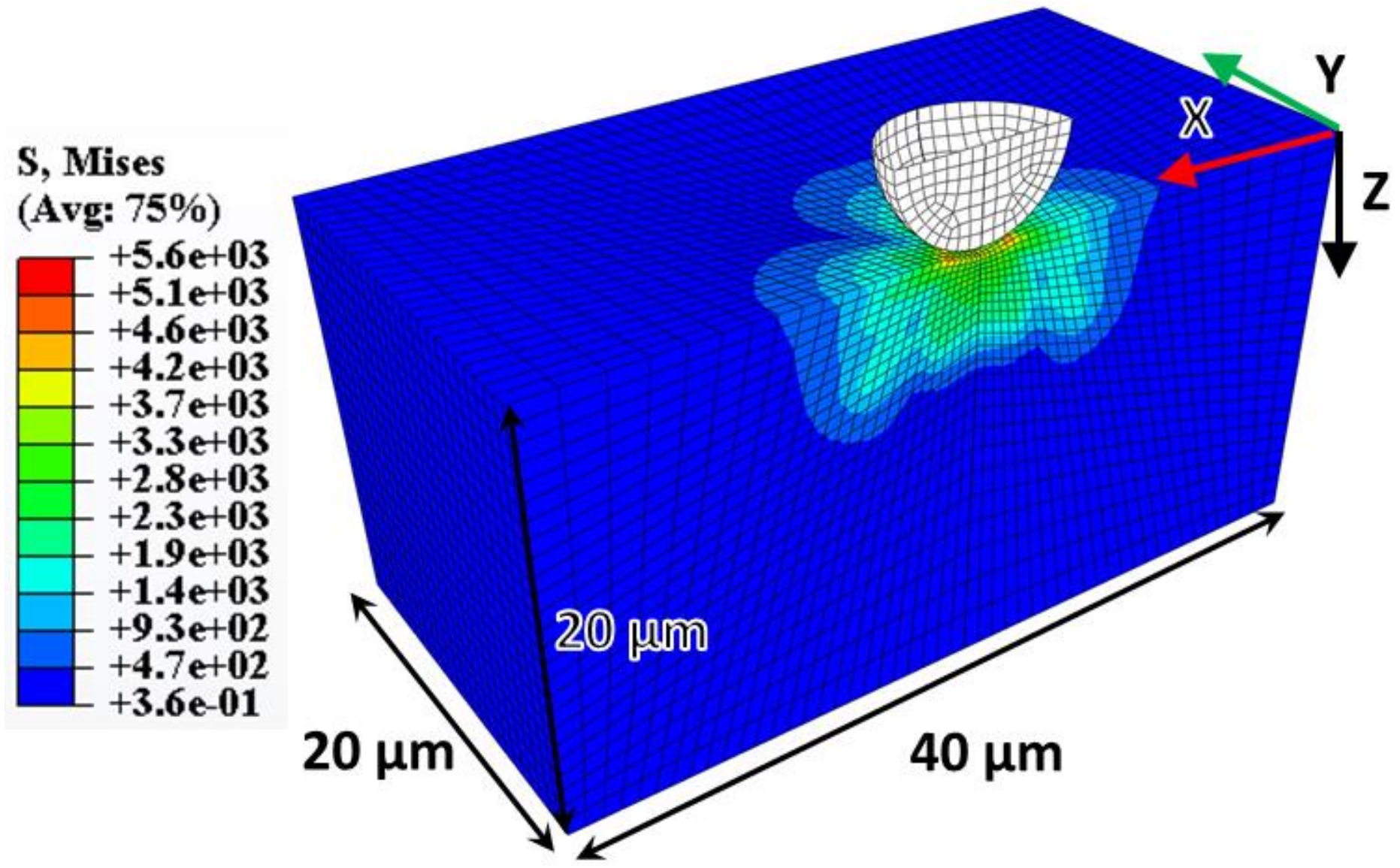

Figure 3 - Von Mises stress representation (after unloading) in the crystal plasticity finite element simulation of a tungsten sample (mirrored at the YZ plane) indented by a 4.2 µm radius spherical indenter. Superimposed are the X, Y, Z coordinate frame and FE mesh used.

Lattice rotations and residual elastic strain fields beneath the indents were extracted from these simulations and were directly compared with the corresponding experimental measurements. Strain gradient crystal plasticity was implemented with a user material subroutine (UMAT) that shares data between gauss points using a common block. The UMAT code was based on the original user element developed by Dunne et al. (Dunne et al., 2007). Further details of the constitutive law and model are provided in Appendix E.

The model was constructed with the initial crystallographic orientation of the sample. Both, experimental and the simulated results are presented in the same sample coordinate frame to enable a direct comparison (Figure 4 and Figure 6).

# 4. Results and Discussion

## 4.1. Residual elastic lattice strains and rotations

Lattice orientation of all sample points, captured by rotation matrix $\boldsymbol{R}$, was measured experimentally by Laue diffraction and also predicted by the CPFE simulations. The average of the rotation matrix, $\boldsymbol{R}$, of points located between 22-25 µm beneath the indent (approximate location of the red dot in Figure 2) was chosen as the reference, $\boldsymbol{R}_{ref}$, and the changes in orientation, $\boldsymbol{R}_{dif}$, of all other points were calculated with respect to $\boldsymbol{R}_{ref}$ (Eq. (44)). $\boldsymbol{R}_{ref}$ captures the combined effect of right handed rotations about the X, Y and Z axis, $\theta_x$, $\theta_y$ and $\theta_z$ respectively. The sequence of rotations is $\theta_x$ first, then $\theta_y$ and finally $\theta_z$ . Given the rotation matrix $\boldsymbol{R}_{dif}$ for every point in the sample, and provided that ${R_{dif}}_{31} \neq \pm 1$, the lattice rotation angles were computed for each sample point using the expressions in Eq. (45) provided in (Slabaugh, 1999):

$$\boldsymbol{R} = \boldsymbol{R}_{dif}\boldsymbol{R}_{ref} \quad (43)$$

$$\boldsymbol{R}_{dif} = \boldsymbol{R}\boldsymbol{R}_{ref}^{-1} \quad (44)$$

$$\theta_x = \tan^{-1}\left(\frac{R_{dif_{32}}}{R_{dif_{33}}}\right), \quad \theta_y = -\sin^{-1}\left(R_{dif_{31}}\right), \qquad \theta_z = \tan^{-1}\left(\frac{R_{dif_{21}}}{R_{dif_{11}}}\right) \tag{45}$$

Figure 4 shows the lattice rotations predicted by CPFE and those measured experimentally, plotted on sections in the YZ plane at different position along the X-axis. Appendix F shows a schematic representation of the nano-indentation process and depicts the lattice rotations expected due to indentation. The lattice rotation directions we observe (Figure 4) agree very well with this. Kysar et al. (Kysar et al., 2010) used electron back-scatter diffraction (EBSD) to measure lattice rotations (spatial resolution of ~3 µm) in a single crystal nickel, indented with a wedge indenter. However, in their study, a two-dimensional deformation state was purposely introduced to eliminate all out-of-plane deformation gradients (within experimental error), such that the resultant dislocation tensor had only two non-zero components. Using the advanced technique of Laue diffraction, we have been able to measure the out-of-plane components of the lattice distortions and therefore the elastic portion of the deformation gradient with sub-micron resolution.

As seen in Figure 4, CPFE predictions match well with the experimental results, for all three rotation components, except at the indent centre (Figure 4, slice 2), where a rapid variation of lattice rotations is seen. A quantitative comparison between the CPFE and the experimental results are made in Figure 5 where line plots corresponding to the contour plots in Figure 4 (b) and (d), have been extracted at a depth of 5 µm beneath the indent (shown by white dotted lines in Figure 4 (a) and (c)). The results for both CPFE calculations and Laue measurements are superimposed. Agreement is quite good, particularly for slices 1 and 3, i.e. for the slices 5 µm either side of the indent. At the indent centre significant discrepancies are visible, likely the result of steep strain gradients which cannot be captured properly due to limited spatial resolution in our measurements. A similar effect, although at a lower spatial resolution (> 2µm), of strong discontinuities in the lattice rotation fields below indents in fcc single crystal nickel, were made by Dahlberg et al. (Dahlberg et al., 2014) using 2D CPFE and EBSD measurements.

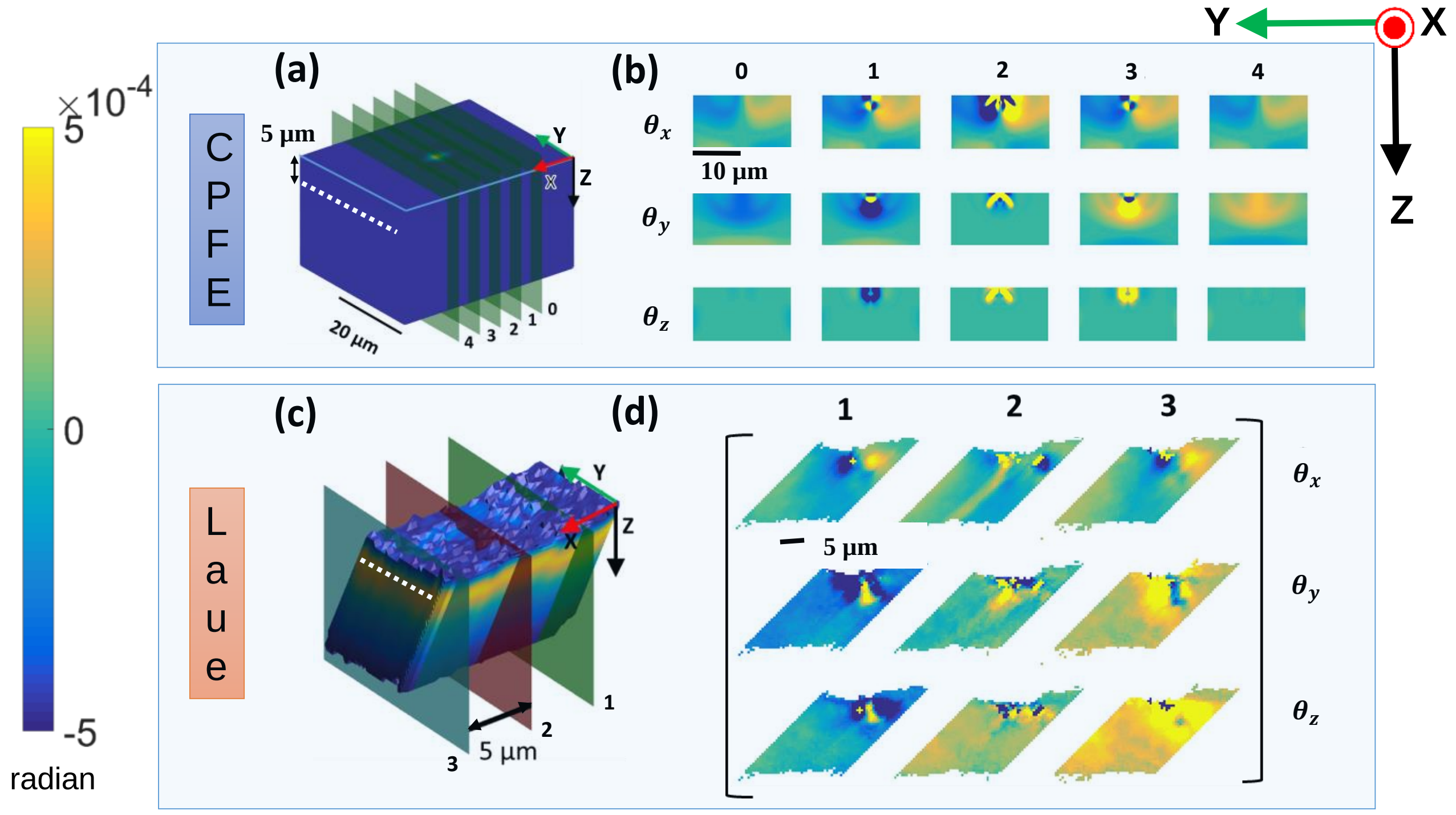

Figure 4 – **CPFE data**: (a) 3D rendering of the simulated volume coloured according to the predicted displacement magnitude along the Z axis of the indented tungsten block. X, Y and Z axes are superimposed. Slices 0-4, drawn on the block, represent the five sections along the X-axis (slice 2 being the indent centre), on which the lattice rotations predicted by CPFE are plotted in (b). **Laue experimental data**: (c) Visualization of the measured sample volume, coloured according to the experimentally measured intensity. Superimposed are the X, Y, and Z axes, as well as the slices on which the measured lattice rotations in (d) are plotted. With respect to the initial crystallographic coordinates, the X axis points in [1 -1 0] direction, the Y in [1 1 0] direction and the Z in the [001] direction.  Slices 1-3 in experiments and simulations are at the same spatial positions and data in (b) and (d) are displayed on the same length- and colour-scale. The dotted white lines through (a) and (c) represent the depth at which the line plots in Figure 5 were extracted.

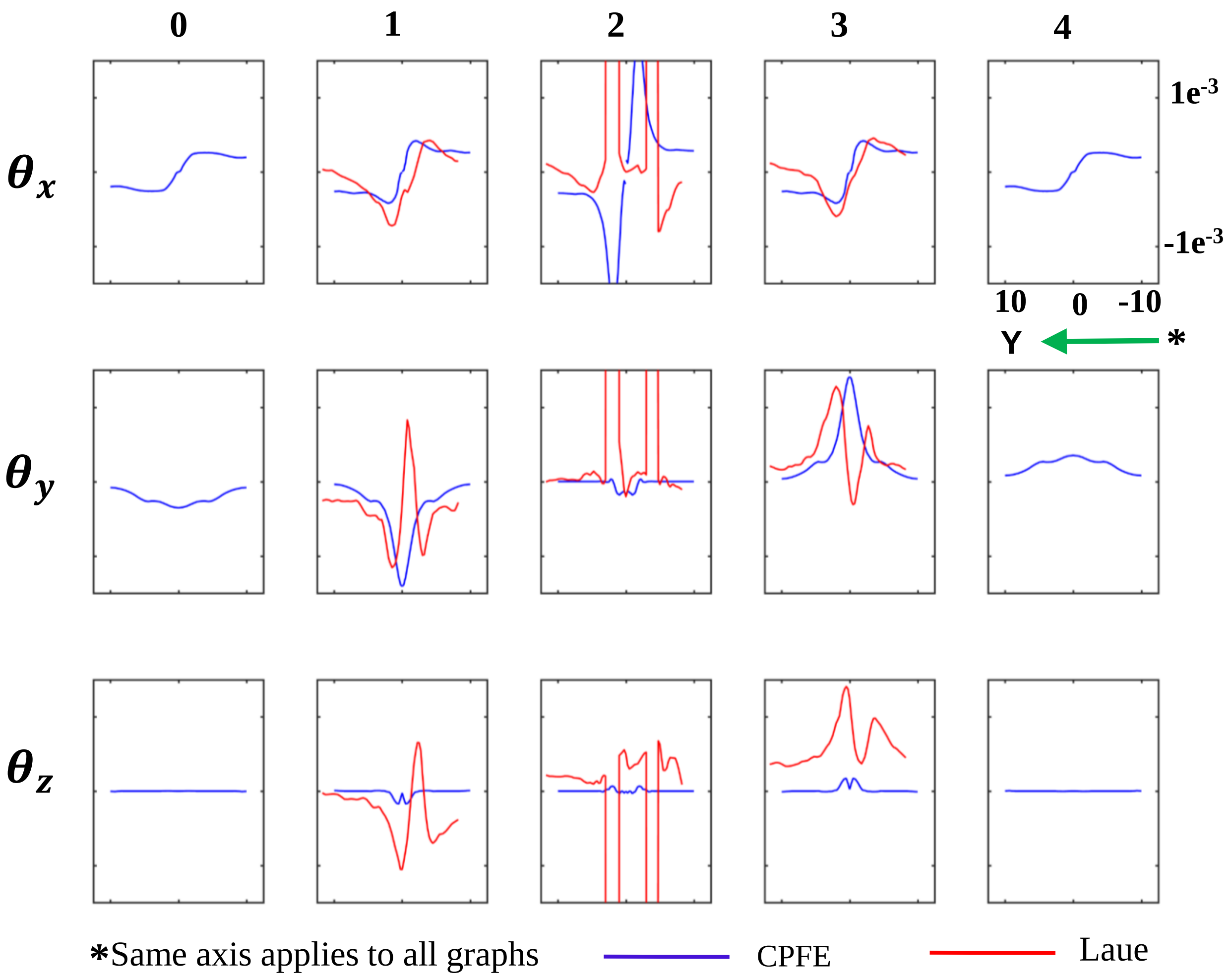

Figure 5 – Line plots corresponding to the contour plots in Figure 4 (b) and (d). Lattice rotations are plotted along a horizontal line 5 µm below the indent (line shown by dotted white lines in Figure 4 (a) and (c)). The slices (0-4) represent the five sections along the X-axis (slice 2 being the indent centre), shown by the YZ planes drawn in Figure 4 (a) and (c).

The experimental Laue measurements only provide the deviatoric lattice strain tensor ($\boldsymbol{\varepsilon}^e_{dev}$), which is related to the total strain tensor ($\boldsymbol{\varepsilon}^e$) by

$$\boldsymbol{\varepsilon}^e = \boldsymbol{\varepsilon}^e_{dev} + \boldsymbol{\varepsilon}^e_{vol} = \boldsymbol{\varepsilon}^e_{dev} + 1/3 \ \mathrm{Tr}\,(\boldsymbol{\varepsilon}^e)\boldsymbol{I} \tag{46}$$

To make a direct comparison with the experimental measurements, the dilatational and plastic strains were removed from the total strain predictions from CPFE. Figure 6 shows the direct components of $\boldsymbol{\varepsilon}^e_{dev}$ predicted by CPFE and as well as the experimentally measured strains, plotted on the same YZ sections at different positions along the X-axis. Qualitatively there is quite good agreement, especially for the $\varepsilon_{zz}$ out-of-plane strain component. A quantitative comparison between the results is made through comparing line plots (Figure 7), corresponding to the contour plots in Figure 6 (b) and (d), extracted at depth of 5 µm beneath the indent (white dotted lines in Figure 6 (a) and (c)). The agreement of the lattice strains is not as clear as for the lattice rotations, but similar features can be identified in the measured and predicted profiles. This is especially the case for slices 1 and 3, i.e. slightly (5 µm away) away from the indent centre.

Figure 8 shows all the components of the symmetric deviatoric strain tensor plotted on XZ and XY planes through the indent centre. Apart from the shear components, $\varepsilon_{xz}$ and $\varepsilon_{yz}$, the strains predicted by CPFE and those measured experimentally agree quite well. In particular the symmetry of the deformation fields is reproduced. Experimental data for the $\varepsilon_{xz}$ and $\varepsilon_{yz}$ components is noisy as the experimental configuration is relatively insensitive to these strain components (Hofmann et al., 2013).

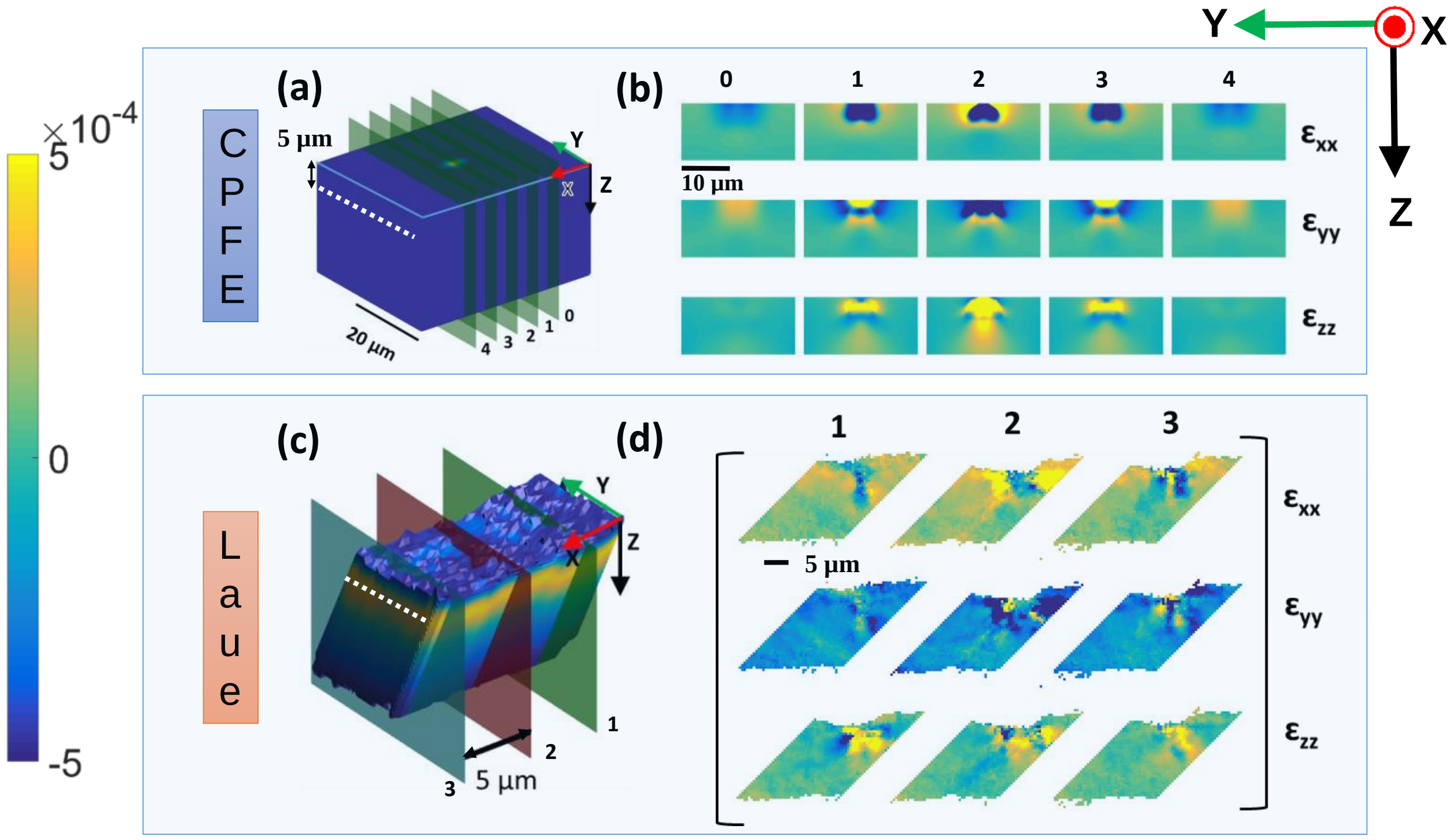

Figure 6 - **For the CPFE data**: (a) The slices 0-4 drawn on the block represent the five sections along the X-axis (slice 2 being at the indent centre) on which the residual deviatoric elastic strains predicted by CPFE are shown in (b). **For the Laue data**: Visualisation of the measured sample volume. Superimposed are the X, Y, and Z axes, as well as the slices on which the measured deviatoric elastic lattice strains in (d) are plotted. With respect to the initial crystallographic coordinates, the X axis points in [1 -1 0] direction, the Y in [1 1 0] direction and the Z in the [001] direction. Slices 1-3 in experiments and simulations are at the same spatial positions and data in (b) and (d) is displayed on the same length- and colour-scales. The dotted white lines through (a) and (c) represent the depth at which the line plots in Figure 7 were extracted.

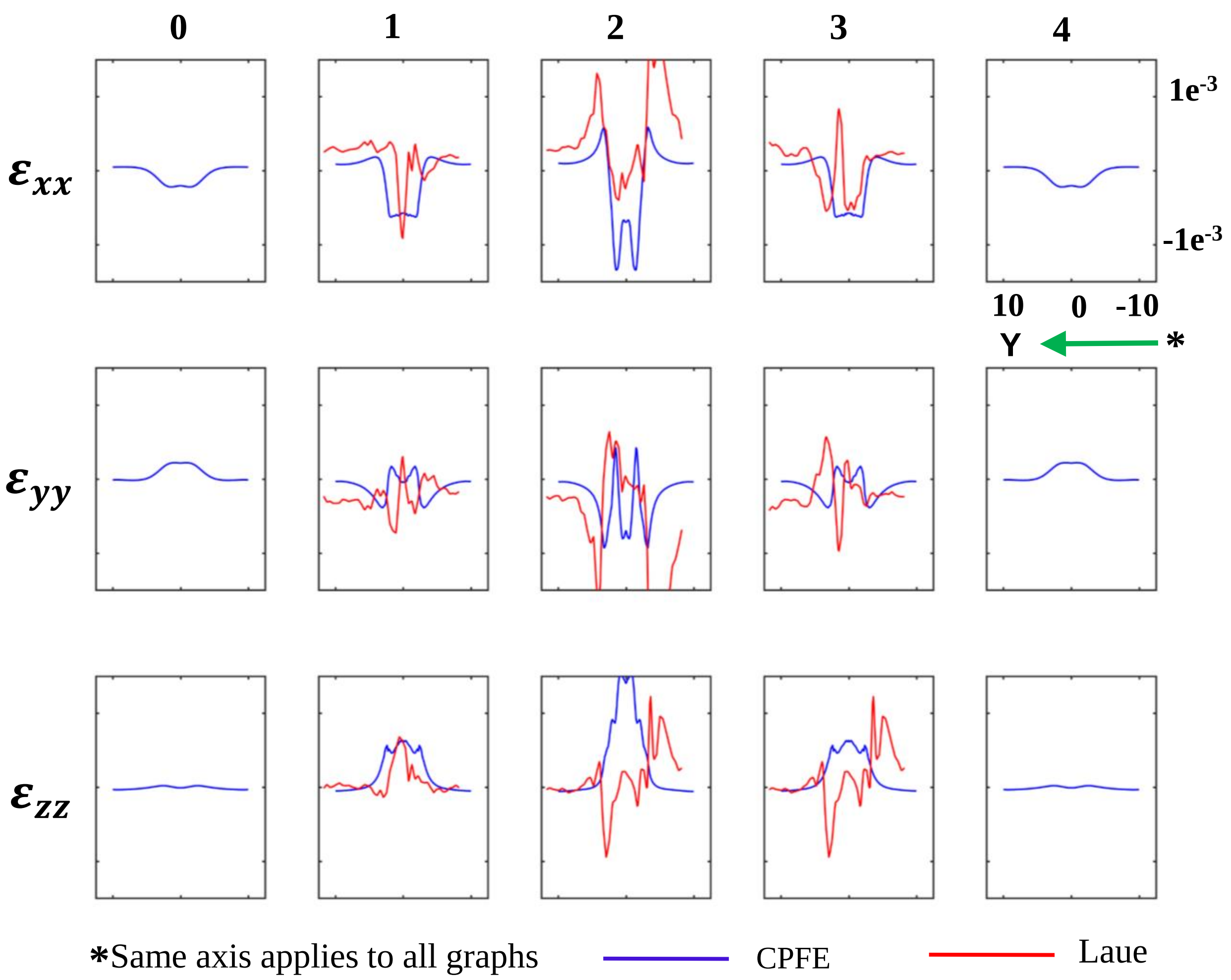

Figure 7 - Line plots corresponding to contour plots in Figure 6 (b) and (d) extracted at a depth of 5 µm below the indent (dotted white lines in Figure 6 (a) and (c)). The slices (0-4) represent the five sections along the X-axis (slice 2 being the indent centre), shown by the YZ planes drawn in Figure 6 (a) and (c).

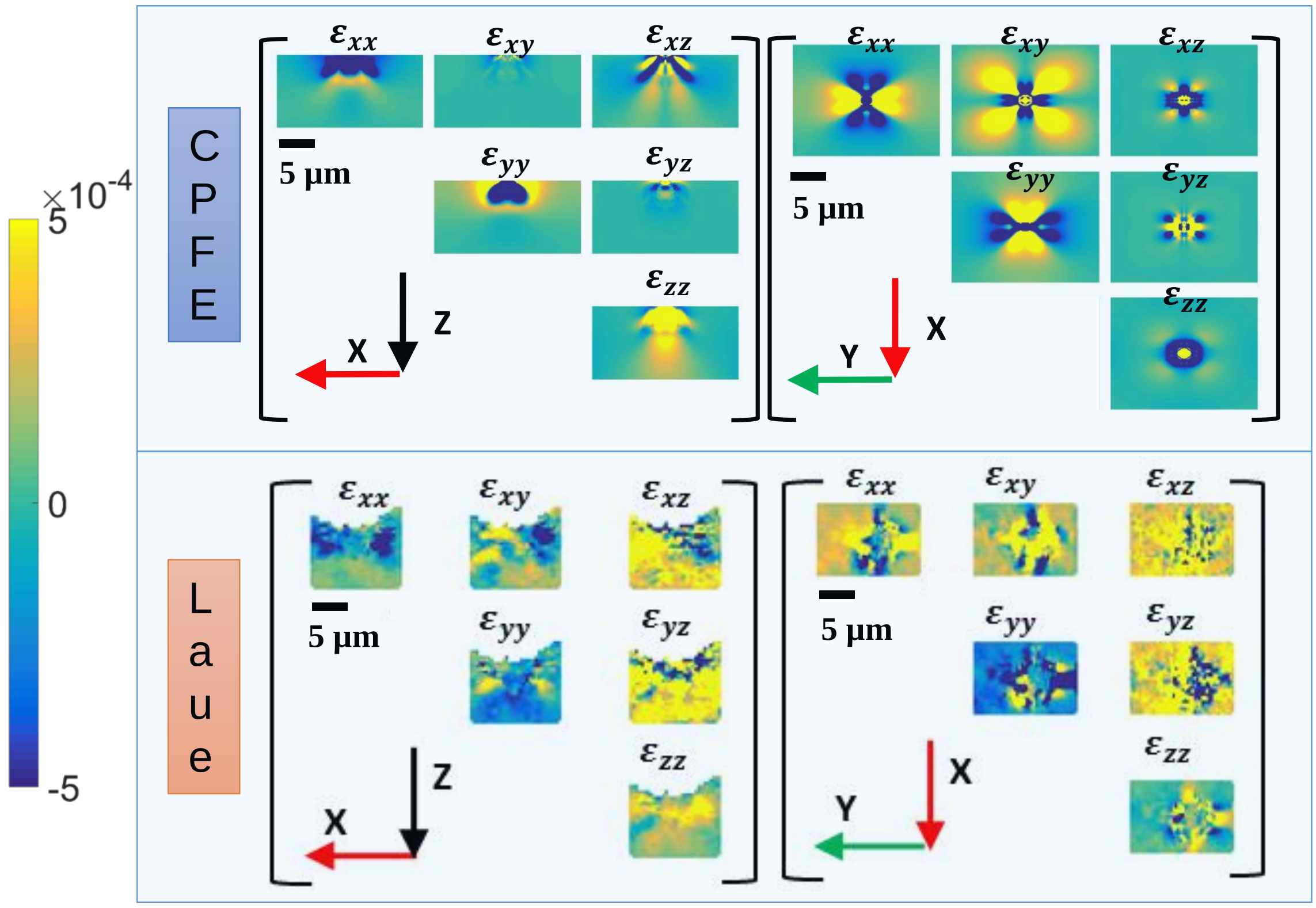

Figure 8 – Residual deviatoric elastic lattice strains predicted by CPFE simulation and experimentally measured by Laue-diffraction. The strains are plotted on sections through the indent centre in the XZ plane and at the indented free surface, the XY plane. Numerical predictions and experimental measurements are shown using the same length- and colour-scales.

Several recent studies have used high-resolution electron backscatter diffraction (HR-EBSD) or high-resolution digital image correlation (HR-DIC) in conjunction with CPFE simulations, to understand strain fields in crystals. For example, Guan et al. (Guan et al., 2017) used HR-DIC and CPFE simulations to investigate the development of strain fields and strain localization in single crystal and oligocrystal nickel subjected to three-point beam bending with cyclic loading. Measurements were restricted to a two-dimensional area on the sample surface and qualitative agreement of only the three in-plane strain components in the XY plane were obtained. Kartal et al (Kartal et al., 2015) used HR-EBSD, on the free surface of a nickel sample, to extract the full residual elastic strain tensor resulting from differences in thermal expansivities between the nickel matrix and a carbide particle embedded within it. A direct comparison of the deviatoric strain measurements from HR-EBSD and CPFE simulations only showed good agreement for the shear component $\varepsilon_{xy}$. In a similar study, Zhang et al. (Zhang et al., 2016), used HR-DIC, HR-EBSD and CPFE simulations to assess the residual strain fields in a polycrystal nickel alloy embedded with a non-metallic agglomerate. They too only find qualitative agreement of the in-plane strains.

The key limitation of HR-EBSD and HR-DIC is their lack of depth-resolved information, thereby allowing examination of the deformation field only at the sample surface. In contrast, micro-beam Laue measurements allows 3D – resolved strain measurement with very good sensitivity of $\sim 10^{-4}$. However, its spatial resolution (~0.5 to 1 µm in 3D) is lower than HR-EBSD (~0.05 µm). Our measurements show surprisingly good agreement between the measured lattice rotations and strains, and those predicted by CPFE (Figure 4 and Figure 6), inspiring some confidence in the use of this combination of techniques for analysing crystal scale deformation.

## 4.2. Dislocation tensor computation

Next we compare the dislocation tensor, $\boldsymbol{\alpha}$, found from Laue diffraction experiments with that predicted by CPFE calculations. Laue diffraction only measures the deviatoric residual elastic strain ($\boldsymbol{\varepsilon}^{e}_{dev}$), while the CPFE simulations provide the full residual strain tensor (both elastic and plastic components). To allow a direct comparison, $\boldsymbol{\alpha}$ was computed from CPFE and experiments using only $\boldsymbol{\varepsilon}^e_{dev}$. The elastic component of the displacement gradient ($\boldsymbol{\beta}^e_{dev}$) was calculated using $\boldsymbol{\varepsilon}^e_{dev}$ and lattice rotation ($\boldsymbol{\omega}^e$) measurements:

$$\boldsymbol{\beta}^e_{dev} = \boldsymbol{\omega}^e + \boldsymbol{\varepsilon}^e_{dev} \tag{47}$$

$\boldsymbol{\alpha}$ was then found by taking the curl of $\boldsymbol{\beta}^e_{dev}$.

Figure 9 shows the dislocation tensor, calculated from CPFE and experimental measurements, plotted on YZ, XZ and XY sections, through the indent centre. From Eq. (32), $\boldsymbol{\alpha} = (\text{curl}\,(\boldsymbol{\beta}^p))^T \approx -(\text{curl}\,(\,\boldsymbol{\beta}^e))^T$ for small strains. The plots of $(\text{curl}\,(\boldsymbol{\beta}^p))^T$ and $-(\text{curl}\,(\,\boldsymbol{\beta}^e_{dev}))^T$, computed from CPFE simulations, are remarkably similar. The CPFE and experimental measurements, both show large and rapid variations of the dislocation tensor near the indent. However, the details of the components of $\boldsymbol{\alpha}$ clearly are quite different. The limited spatial resolution in experiments is likely the main reason. This is particularly the case since the curl operation takes the gradient of the measured lattice strains and rotations, making it very sensitive to experimental uncertainties, especially in the presence of steep strain gradients. Interestingly the XY plots clearly show the same symmetry in both experiments and simulations.

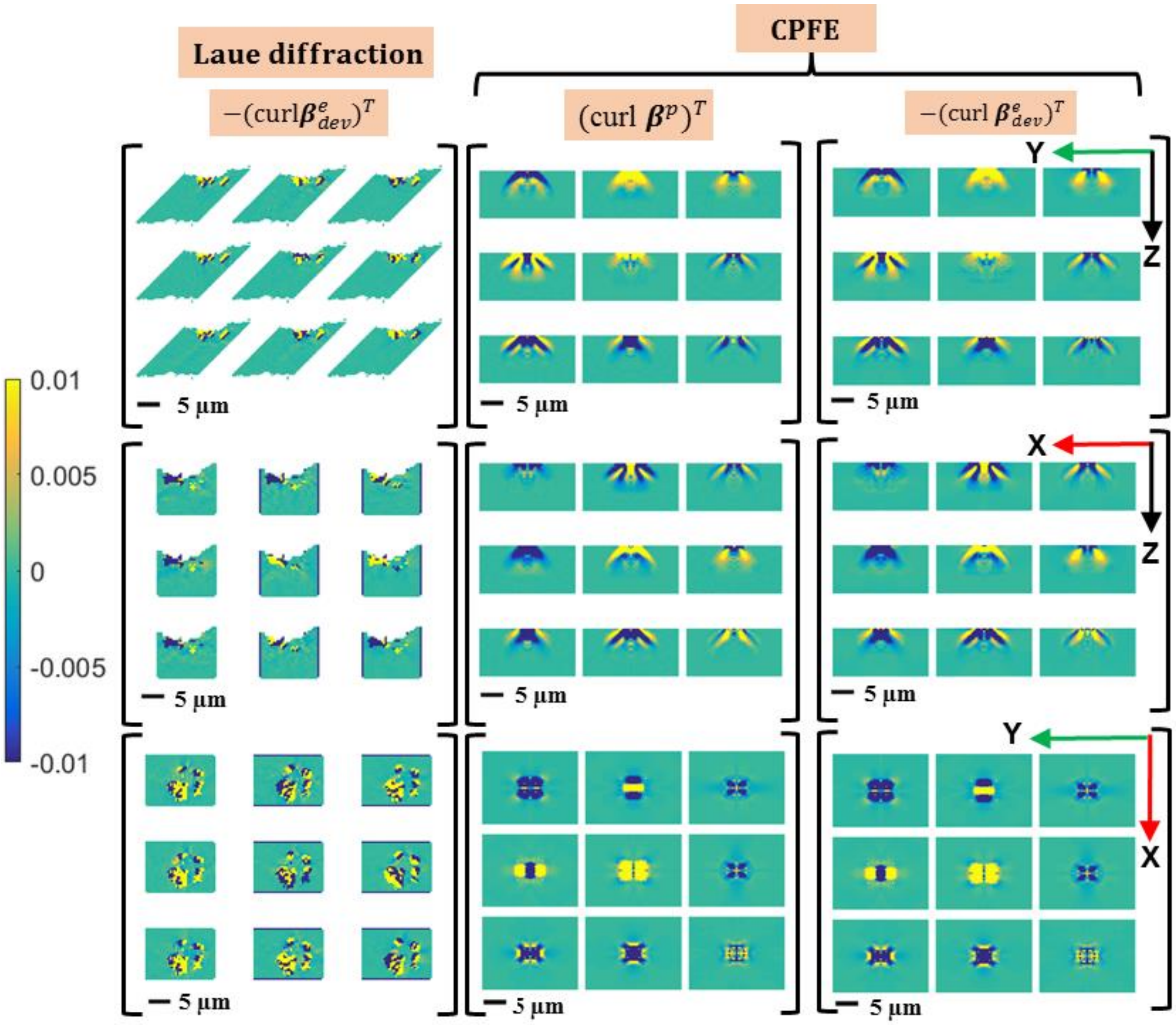

Figure 9 - The dislocation tensor $\boldsymbol{\alpha}$ of the deformation field beneath spherical indents in pure tungsten. Plots are shown on sections thought the indent centre on YZ, XZ and the XY planes. Colour scale has units of 1/µm and $\boldsymbol{\alpha}$ is represented as $\begin{bmatrix} \alpha_{11} & \alpha_{12} & \alpha_{13} \\ \alpha_{21} & \alpha_{22} & \alpha_{23} \\ \alpha_{31} & \alpha_{32} & \alpha_{33} \end{bmatrix}$.

An important question concerns the error incurred by neglecting the volumetric component of the elastic lattice strain. Consider the difference, δ, between $(-\mathrm{curl}\ (\boldsymbol{\beta}^{e}))^{T}$ and $(-\mathrm{curl}\ (\boldsymbol{\beta}^{e}_{dev}))^{T}$, which both provide approximations to $\boldsymbol{\alpha}$:

$$\boldsymbol{\alpha} \cong (-\mathrm{curl}\ (\boldsymbol{\beta}^{e}))^{T} = (-\mathrm{curl}\ (\boldsymbol{\varepsilon}^{e} + \boldsymbol{\omega}^{e}))^{T} = (-\mathrm{curl}\ (\boldsymbol{\varepsilon}^{e}_{dev} + \boldsymbol{\varepsilon}^{e}_{vol} + \boldsymbol{\omega}^{e}))^{T} \tag{48}$$

$$\boldsymbol{\alpha} \cong (-\mathrm{curl}\ (\boldsymbol{\beta}^{e}_{dev}))^{T} = (-\mathrm{curl}\ (\boldsymbol{\varepsilon}^{e}_{dev} + \boldsymbol{\omega}^{e}))^{T} \tag{49}$$

The difference, δ, corresponds to the curl of the volumetric strain component ($\boldsymbol{\varepsilon}^{e}_{vol}$):

$$\delta = \left(-\mathrm{curl}\ (\boldsymbol{\varepsilon}^{e}_{dev} + \boldsymbol{\varepsilon}^{e}_{vol} + \boldsymbol{\omega}^{e})\right)^{T} - \left(-\mathrm{curl}\ (\boldsymbol{\varepsilon}^{e}_{dev} + \boldsymbol{\omega}^{e})\right)^{T} = \left(-\mathrm{curl}\ (\boldsymbol{\varepsilon}^{e}_{vol})\right)^{T} \tag{50}$$

Figure 10 shows plots of δ, computed from the CPFE simulations, on the same YZ, XZ and the XY planes through the indent centre as used in Figure 9. A comparison of Figure 9 (column 3: depicting the components of $\boldsymbol{\alpha}$ calculated using $\left(-\mathrm{curl}\ (\boldsymbol{\beta}^{e}_{dev})\right)^{T}$) and Figure 10 shows that the magnitude of δ is substantially smaller than that of $\boldsymbol{\alpha}$. This suggests that effect of $\left(-\mathrm{curl}\ (\boldsymbol{\varepsilon}^{e}_{vol})\right)^{T}$ on the calculated components of $\boldsymbol{\alpha}$ is small.

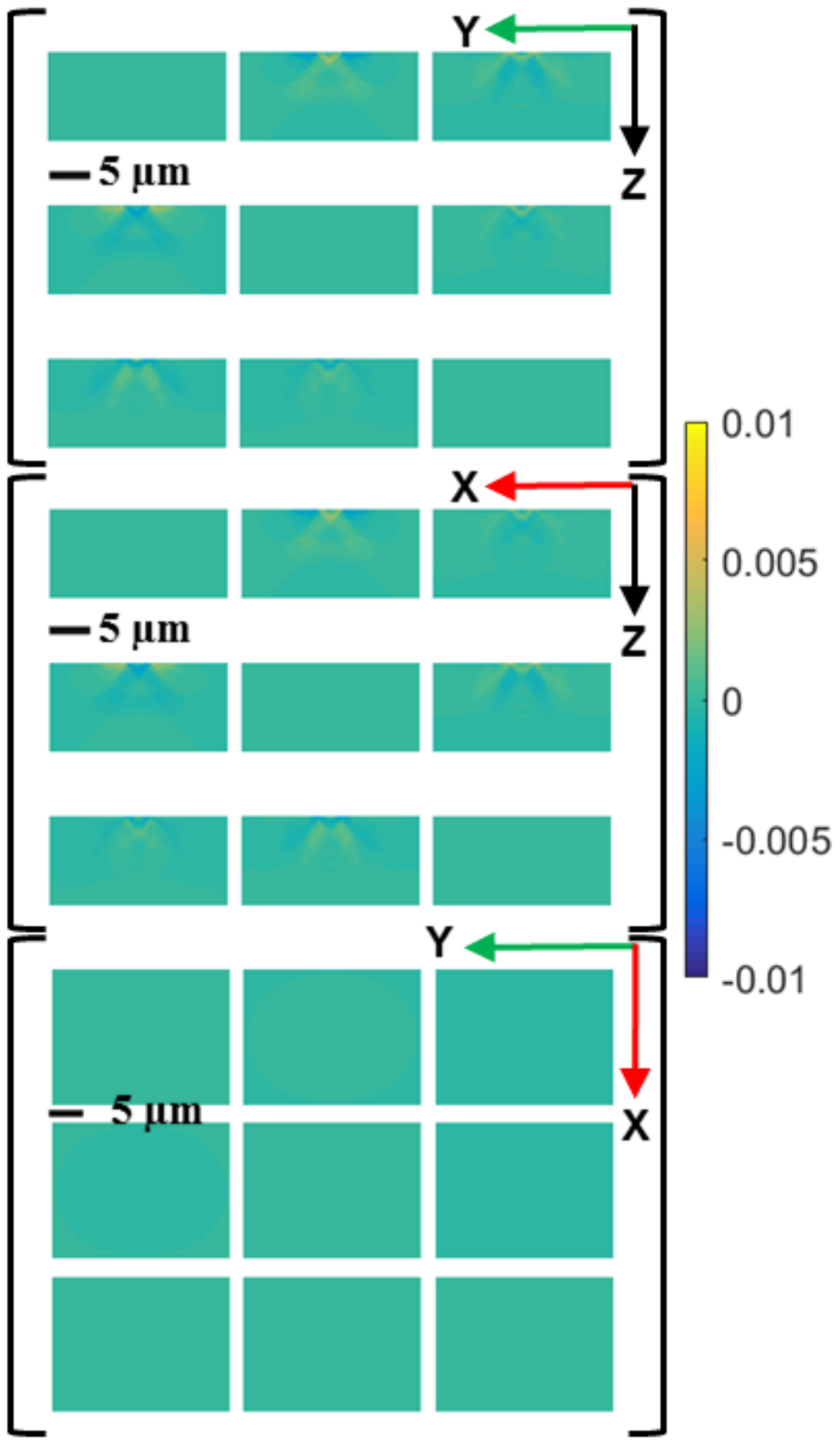

Figure 10 – Plot of $\delta = \left(-\text{curl}\,(\boldsymbol{\beta}^{e})\right)^{T} - \left(-\text{curl}\,(\boldsymbol{\beta}^{e}_{dev})\right)^{T}$ at the indent centre, on the YZ, XZ and XY planes. The colour scale has units of 1/µm and the representation of the components of δ is the same as that of $\boldsymbol{\alpha}$ in Figure 9.

The lack of sensitivity of $\boldsymbol{\alpha}$ to the volumetric part of the elastic strain tensor is an important result since many experimental techniques (e.g. Micro-beam Laue diffraction, HR-EBSD) can only readily measure deviatoric elastic strain. Nye's original formulation [26] only considered lattice rotations, implying that for small deformations the effect of lattice rotations dominates over that of lattice strains. The volumetric component of the elastic strain tensor is expected to play an even smaller part since plastic deformation, accommodated by crystallographic slip, is an isochoric process.

The 3D depth-resolved measurements of deviatoric lattice strain and rotation, possible with micro-beam Laue diffraction, allow determination of all nine components of the dislocation tensor. This is in contrast to surface techniques, such as HR-EBSD (Wilkinson and Randman, 2010), (Wallis et al., 2016), (Jiang et al., 2015), (Ruggles et al., 2016) or micro-Laue diffraction without depth resolution (Irastorza-Landa et al., 2017), where terms of the dislocation tensor Eq. (38) depending on $\frac{\partial}{\partial x_3}$ remain unknown. Hence, without depth-resolution, only three of the nine elements of $\boldsymbol{\alpha}$ can be explicitly determined. If the assumption is made that the effect of lattice strains is negligible, five components may be determined (Pantleon, 2008). This means that GND densities determined from 2D surface methods will always constitute a lower bound estimate.

## 4.3. GND density computation

GND density was computed using $L^2$ and $L^1$ optimisation techniques for both experimental measurements and CPFE simulations. For tungsten we assume that deformation is accommodated by dislocations with a/2<111> Burgers' vector slipping on {110} planes (Marichal et al., 2013; Srivastava et al., 2013) (list of Burgers' vectors and line directions in Appendix G). Furthermore, we assume dislocations to have either pure edge or pure screw character. This results in 16 distinct dislocation types; four screw types with <111> line directions and twelve edge type with <112> line directions.

### 4.3.1 $L^1$ vs $L^2$ Optimisation

The GND densities of all sixteen dislocation types, determined using the $L^2$ optimisation method (Eq. (39) and (41)), are shown in Figure 11 and Figure 13 for experiments and CPFE simulations respectively. In both figures dislocation densities are plotted on YZ, XZ and XY sections through the indent centre. GND densities, determined using the $L^1$ optimisation method (Eq. (39) and (42)), are shown in Figure 12 and Figure 14 for experiments and CPFE simulations respectively. For the $L^2$ optimisation the GND density is distributed (almost evenly) over all slip systems. The reason is that minimisation of $\sum_j \rho_j^2 = \boldsymbol{\rho}^T.\boldsymbol{\rho}$ associates a larger penalty with slip systems that have high dislocation density. Thus a solution where dislocation density is distributed amongst slip systems is favourable. In contrast, the $L^1$ scheme minimises the total energy (weighted line length) namely $(1-\upsilon)^{-1}\sum_j \rho_j^{edge} + \sum_j \rho_j^{screw}$. Here a much greater variation of dislocation density distribution between slip systems is observed.

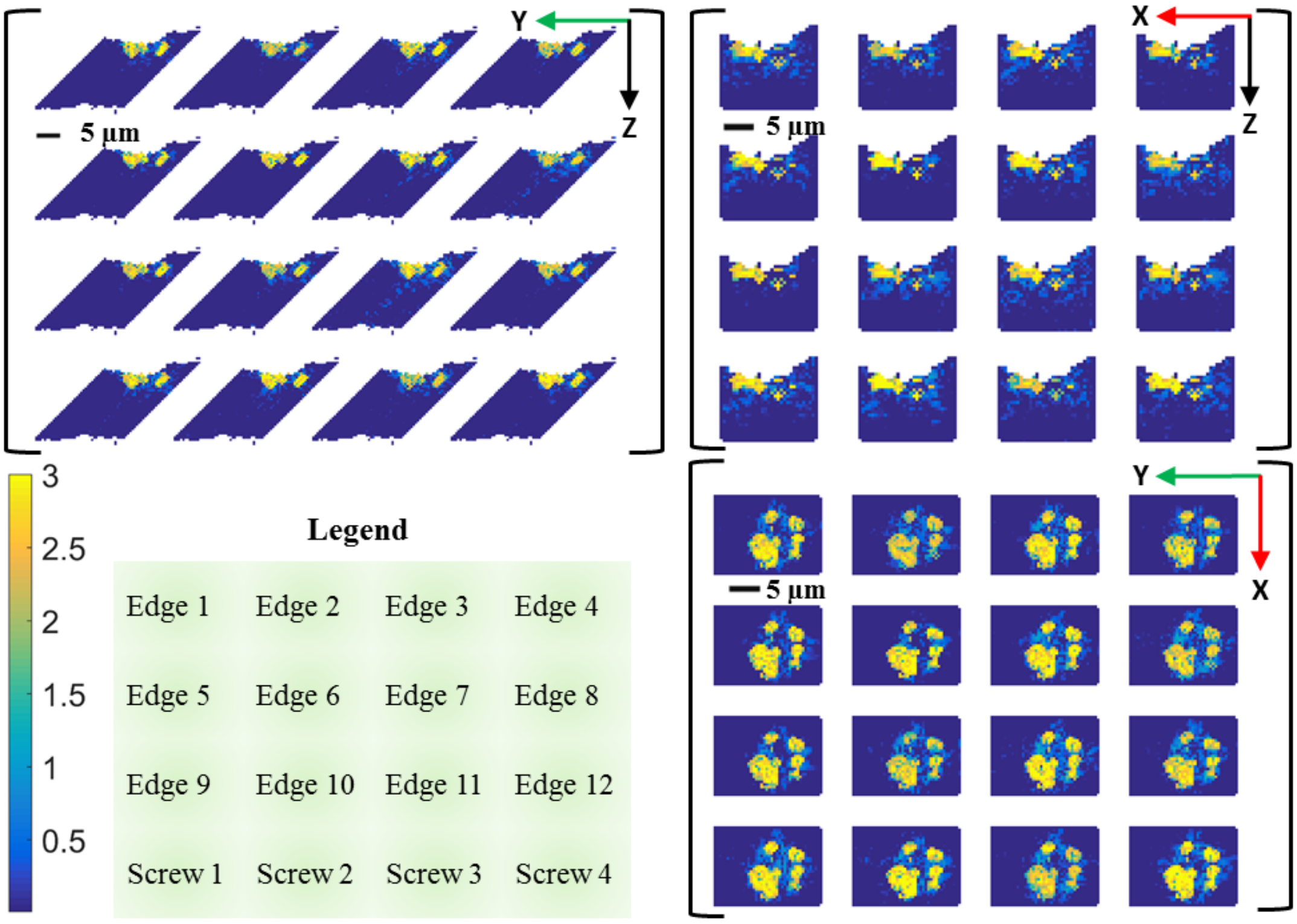

Figure 11 – Experimental dislocation densities obtained by $L^2$ optimization method plotted at the indent centre on the YZ, XZ and the XY plane. Colour scale shows $\log_{10}(\rho)$ with ρ in 1/µm$^2$. Scale bar = 5 µm.

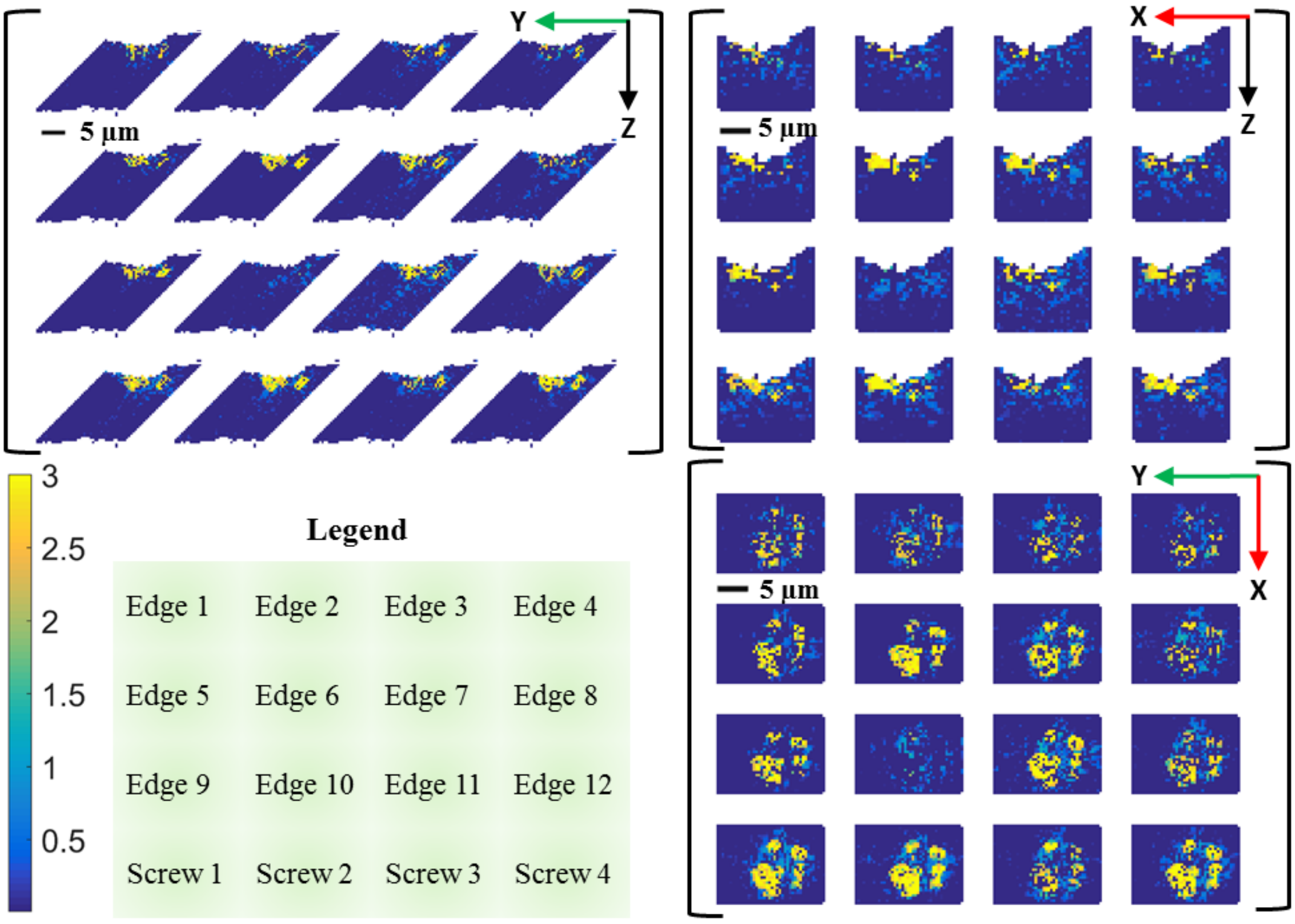

Figure 12 – Experimental dislocation densities obtained by $L^1$ optimization method plotted at the indent centre on the YZ, XZ and the XY plane. Colour scale shows $\log_{10}(\rho)$ with ρ in 1/µm$^2$. Scale bar = 5 µm.

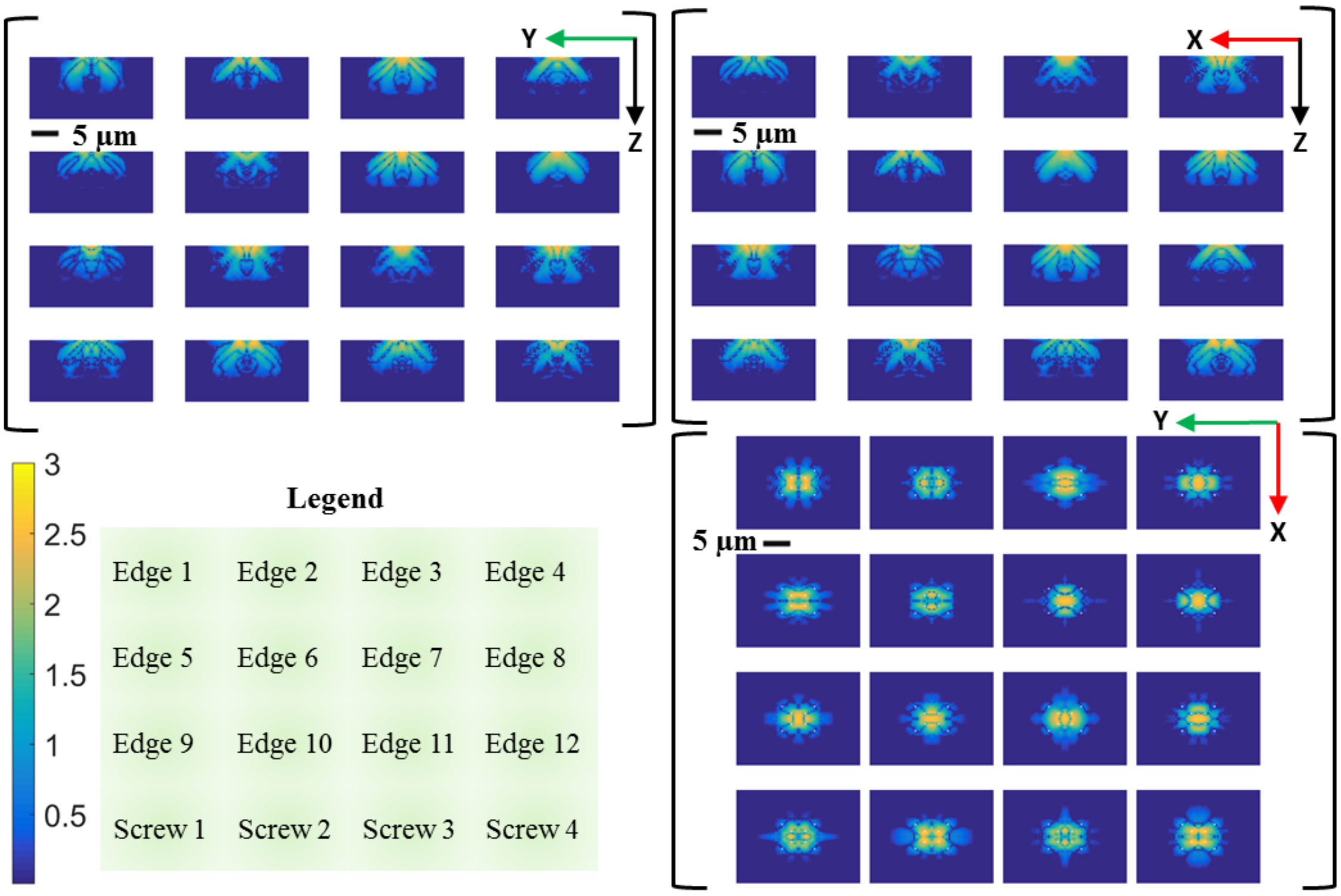

Figure 13 – CPFEM Dislocation densities obtained by $L^2$ optimization method plotted at the indent centre on the YZ, XZ and the XY plane. Colour scale shows $\log_{10}(\rho)$ with ρ in $1/\mu m^2$. Scale bar = 5 µm.

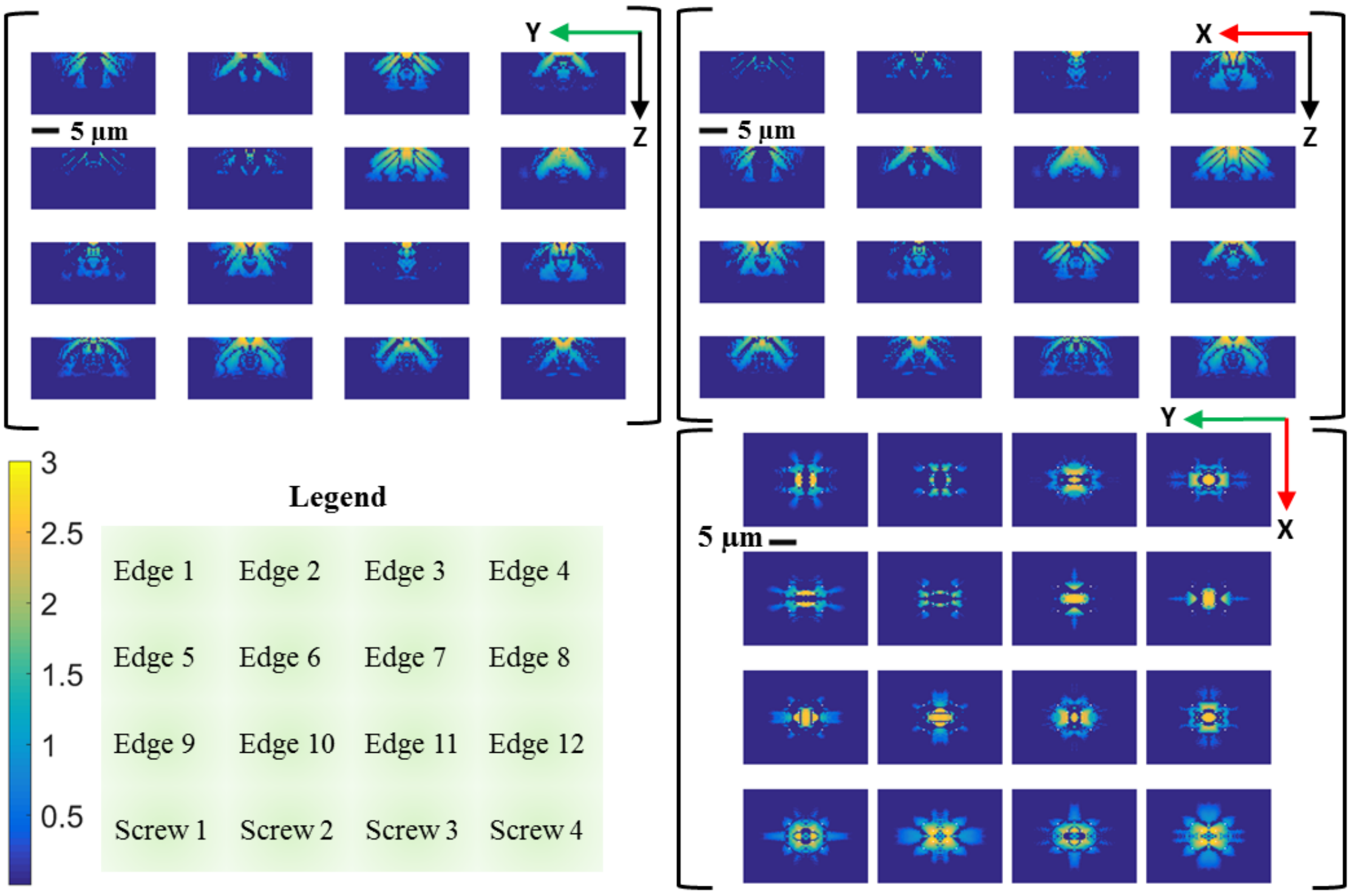

Figure 14 – CPFEM Dislocation densities obtained by $L^1$ optimization method plotted at the indent centre on the YZ, XZ and the XY plane. Colour scale shows $\log_{10}(\rho)$ with ρ in $1/\mu m^2$. Scale bar = 5 µm.

Arsenlis and Parks (Arsenlis and Parks, 1999) compared $L^1$ and $L^2$ optimisation techniques for an fcc crystal and found that $L^1$ produced more accurate results for a dislocation structure consisting of two dislocation lines. In contrast, the $L^2$ method predicted complex dislocation structures with multiple dislocation lines. Randman et al. (Wilkinson and Randman, 2010) and Ruggles et al. (Ruggles et al., 2016) also observed that $L^1$ minimisation generates an uneven distribution of GND density over individual slip systems. However, in their study, no corresponding comparison was made to densities obtained using the $L^2$ method. Our direct comparison of $L^1$ and $L^2$ methods, for both experimental and CPFE datasets, is consistent with these observations. It highlights that $L^2$ optimisation leads to an unphysical spreading of dislocation density over many slip systems, making the use of the $L^1$ method for accurate estimation of GND densities on individual slip systems essential. This effect will be of particular importance for crystal plasticity simulations where a distinction between the cutting density and mobile density associated with particular slip system is made (Roters et al., 2010).

The total dislocation density (i.e. summed over all slip systems) computed through both methods is remarkably similar. This is shown in Figure 15 where the total GND densities (calculated using $L^1$ and $L^2$ optimisations) from experiments and CPFE are plotted on YZ, XZ and XY sections through the indent centre. Thus, if only the total dislocation density in the sample is required, either of the optimisation techniques may be used. In this case, the $L^2$ optimisation, which is far more straightforward to implement and is computationally cheaper, is preferable.

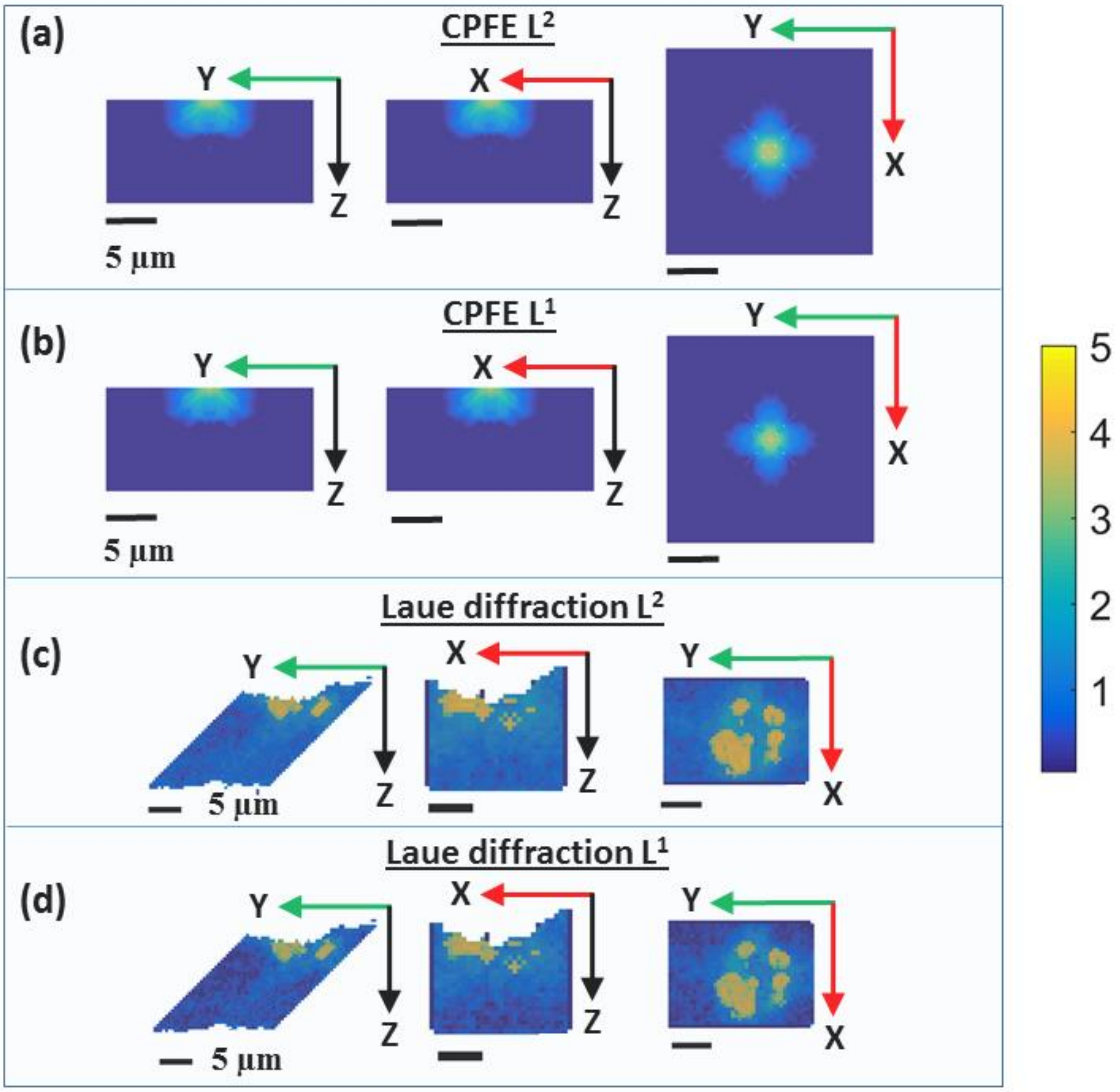

Figure 15 - Total dislocation density plotted on YZ, XZ and XY sections through the indent centre. The GND densities predicted by $L^2$ and $L^1$ optimisation based on CPFE data are shown in (a) and (b) respectively. $L^2$ and $L^1$ results from the experimental measurements are plotted in (c) and (d) respectively. The colour scale shows $\log_{10}(\rho)$ with $\rho$ in lines/$\mu m^2$.

# 5. Conclusions

The dislocation tensor captures the dislocation population required to accommodate inhomogeneous plastic deformation. It is linked to the density of geometrically necessary dislocations (GNDs), and can be equated to lattice deformation in the form of lattice rotations and lattice strains. This provides a very useful relationship between GND density and the lattice curvature it accommodates. Here we have provided a comprehensive review of the theoretical background of the computation of Nye's dislocation tensor and the underlying GND density in plastically deformed materials. Comparing CPFE simulations and X-ray diffraction measurements of lattice distortions associated with spherical nano-indents in tungsten, a number of important conclusions can be reached:

- The relationship between different curl definitions, used to compute the dislocation tensor, has been explored. Table 1 provides a summary of the dislocation tensor in terms of both elastic and plastic deformation gradients, for cases of small and large deformations. Importantly the different curl definitions in use, if applied consistently, all lead to the same result.
- Lattice rotations and lattice strains beneath a spherical nano-indent in a tungsten single crystal are considered. CPFE and synchrotron X-ray micro-beam Laue measurements show good qualitative agreement, particularly for the 3D distribution of lattice rotations.
- The contribution of $\left(-\mathrm{curl}\,(\boldsymbol{\varepsilon}^{e}_{vol})\right)^{T}$, the curl of the volumetric part of the elastic strain tensor, to the dislocation tensor is small. This is an important result since many experimental strain measurement techniques, such as white-beam Laue diffraction and HR-EBSD, can only measure the deviatoric lattice strain tensor.
- $L^1$ optimisation recovers a more heterogeneous distribution of GND density over individual slip systems. In contrast $L^2$ optimisation distributes GND density almost uniformly over all slip systems. Thus, if GND density on specific slip systems is required, the physically-based $L^1$ minimisation should be used.

- The total GND density determined by either $L^1$ or $L^2$ minimisation is remarkably similar. Thus the computationally simpler $L^2$ optimisation may be used if only total GND density is required.

# Appendix A

Derivation of the three common definitions of the curl of a second order tensor.

**Precurl**

$$\boldsymbol{R} = (\nabla \times \boldsymbol{P}) = \left(\widehat{e_i}\frac{\partial}{\partial x_i}\right) \times P_{jm}\widehat{e_j} \otimes \widehat{e_m} = \left(\widehat{e_i} \times \widehat{e_j}\right) \otimes \widehat{e_m}\frac{\partial P_{jm}}{\partial x_i} = \widehat{e_k} \otimes \widehat{e_m} \in_{kij} P_{jm,i} \quad \text{(A.1)}$$

Thus, $\boldsymbol{R}_{km} = (\nabla \times \boldsymbol{P})_{km} = \in_{ijk} P_{jm,i}$

**Postcurl**

$$\boldsymbol{S} = (\boldsymbol{V} \times \nabla) = V_{kj}\widehat{e_k} \otimes \widehat{e_j} \times \left(\widehat{e_i}\frac{\partial}{\partial x_i}\right) = V_{kj}\, \widehat{e_k} \otimes \left(\widehat{e_j} \times \widehat{e_i}\right)\frac{\partial}{\partial x_i} \quad \text{(A.2)}$$
$$= V_{kj}\, \widehat{e_k} \otimes \widehat{e_m} \in_{mji} \frac{\partial}{\partial x_i} = \in_{mji} V_{kj,i}\, \widehat{e_k} \otimes \widehat{e_m}$$

Thus, $\boldsymbol{S}_{km} = (\boldsymbol{V} \times \nabla)_{km} = \in_{mji} V_{kj,i} = -\in_{ijm} V_{kj,i}$

**Curl3**

If $\boldsymbol{c}$ is a constant vector then using the definition that $\nabla \times (\boldsymbol{c}.\boldsymbol{V}) = (\nabla \times \boldsymbol{V}).\boldsymbol{c}$ and recalling that

$$\boldsymbol{v} = \boldsymbol{c}.\boldsymbol{V} = c_m V_{mj}\widehat{e_j} \quad \text{(A.3)}$$

we can formulate the curl of a matrix using the familiar definition of the curl of a vector

$$(\nabla \times \boldsymbol{v})_k = \left(\widehat{e_i}\frac{\partial}{\partial x_i}\right) \times c_m V_{mj}\widehat{e_j} = \left(\widehat{e_i} \times \widehat{e_j}\right)c_m\frac{\partial V_{mj}}{\partial x_i} = \in_{kij} c_m V_{mj,i}\widehat{e_k}$$
$$= (\in_{ijk} V_{mj,i}\widehat{e_k} \otimes \widehat{e_m}).\boldsymbol{c} = (\nabla \times \boldsymbol{V}).\boldsymbol{c}$$

Thus, $\boldsymbol{Q}_{km} = (\nabla \times \boldsymbol{V})_{km} = \in_{ijk} V_{mj,i}$

# Appendix B

Let $\boldsymbol{P}$ be defined as a second order tensor

$$\boldsymbol{P} = \begin{bmatrix} a_1 & a_2 & a_3 \\ b_1 & b_2 & b_3 \\ c_1 & c_2 & c_3 \end{bmatrix} \tag{B.1}$$

<u>Proof of curl3 $(\boldsymbol{V}) = \text{Precurl}\left(\boldsymbol{V}^{\mathrm{T}}\right)$</u>

Given the definition of $\boldsymbol{P}$ in Eq. (B.1), as per the above postulated statement, $\boldsymbol{V}$ will then be defined as

$$\boldsymbol{V} = \boldsymbol{P}^{\mathrm{T}} = \begin{bmatrix} a_1 & b_1 & c_1 \\ a_2 & b_2 & c_2 \\ a_3 & b_3 & c_3 \end{bmatrix} \tag{B.2}$$

Now, using the "curl3" definition, $Q_{km}$, can be written explicitly as

$$Q_{km} = (\nabla \times \boldsymbol{V})_{km} = \epsilon_{ijk} V_{mj,i} = \begin{bmatrix} \frac{\partial V_{13}}{\partial x_2} - \frac{\partial V_{12}}{\partial x_3} & \frac{\partial V_{23}}{\partial x_2} - \frac{\partial V_{22}}{\partial x_3} & \frac{\partial V_{33}}{\partial x_2} - \frac{\partial V_{32}}{\partial x_3} \\ \frac{\partial V_{11}}{\partial x_3} - \frac{\partial V_{13}}{\partial x_1} & \frac{\partial V_{21}}{\partial x_3} - \frac{\partial V_{23}}{\partial x_1} & \frac{\partial V_{31}}{\partial x_3} - \frac{\partial V_{33}}{\partial x_1} \\ \frac{\partial V_{12}}{\partial x_1} - \frac{\partial V_{11}}{\partial x_2} & \frac{\partial V_{22}}{\partial x_1} - \frac{\partial V_{21}}{\partial x_2} & \frac{\partial V_{32}}{\partial x_1} - \frac{\partial V_{31}}{\partial x_2} \end{bmatrix} \tag{B.3}$$

Substituting the defined components of $\boldsymbol{V}$ (Eq. (B.2)) in Eq. (B.3), gives

$$Q_{km} = \begin{bmatrix} \frac{\partial c_1}{\partial x_2} - \frac{\partial b_1}{\partial x_3} & \frac{\partial c_2}{\partial x_2} - \frac{\partial b_2}{\partial x_3} & \frac{\partial c_3}{\partial x_2} - \frac{\partial b_3}{\partial x_3} \\ \frac{\partial a_1}{\partial x_3} - \frac{\partial c_1}{\partial x_1} & \frac{\partial a_2}{\partial x_3} - \frac{\partial c_2}{\partial x_1} & \frac{\partial a_3}{\partial x_3} - \frac{\partial c_3}{\partial x_1} \\ \frac{\partial b_1}{\partial x_1} - \frac{\partial a_1}{\partial x_2} & \frac{\partial b_2}{\partial x_1} - \frac{\partial a_2}{\partial x_2} & \frac{\partial b_3}{\partial x_1} - \frac{\partial a_3}{\partial x_2} \end{bmatrix} \tag{B.4}$$

Likewise, the precurl definition may be explicitly written as

$$R_{km} = (\nabla \times \boldsymbol{P})_{km} = \epsilon_{ijk} P_{jm,i} = \begin{bmatrix} \frac{\partial P_{31}}{\partial x_2} - \frac{\partial P_{21}}{\partial x_3} & \frac{\partial P_{32}}{\partial x_2} - \frac{\partial P_{22}}{\partial x_3} & \frac{\partial P_{33}}{\partial x_2} - \frac{\partial P_{23}}{\partial x_3} \\ \frac{\partial P_{11}}{\partial x_3} - \frac{\partial P_{31}}{\partial x_1} & \frac{\partial P_{12}}{\partial x_3} - \frac{\partial P_{32}}{\partial x_1} & \frac{\partial P_{13}}{\partial x_3} - \frac{\partial P_{33}}{\partial x_1} \\ \frac{\partial P_{21}}{\partial x_1} - \frac{\partial P_{11}}{\partial x_2} & \frac{\partial P_{22}}{\partial x_1} - \frac{\partial P_{12}}{\partial x_2} & \frac{\partial P_{23}}{\partial x_1} - \frac{\partial P_{13}}{\partial x_2} \end{bmatrix} \quad \text{(B.5)}$$

By substituting the components of $\boldsymbol{P}$ from Eq. (B.1), $R_{km}$ may be re-stated as

$$R_{km} = \begin{bmatrix} \frac{\partial c_1}{\partial x_2} - \frac{\partial b_1}{\partial x_3} & \frac{\partial c_2}{\partial x_2} - \frac{\partial b_2}{\partial x_3} & \frac{\partial c_3}{\partial x_2} - \frac{\partial b_3}{\partial x_3} \\ \frac{\partial a_1}{\partial x_3} - \frac{\partial c_1}{\partial x_1} & \frac{\partial a_2}{\partial x_3} - \frac{\partial c_2}{\partial x_1} & \frac{\partial a_3}{\partial x_3} - \frac{\partial c_3}{\partial x_1} \\ \frac{\partial b_1}{\partial x_1} - \frac{\partial a_1}{\partial x_2} & \frac{\partial b_2}{\partial x_1} - \frac{\partial a_2}{\partial x_2} & \frac{\partial b_3}{\partial x_1} - \frac{\partial a_3}{\partial x_2} \end{bmatrix} \quad \text{(B.6)}$$

Comparing Eq. (B.4) and Eq. (B.6), it is seen that $Q_{km} = R_{km}$, thus proving that precurl of $\boldsymbol{V}$ is equal to the curl of $\boldsymbol{P}$, provided $\boldsymbol{V} = \boldsymbol{P}^{\mathrm{T}}$.

Proof of curl3($\boldsymbol{V}$) = $-\left(\text{Postcurl }(\boldsymbol{V})\right)^{\mathrm{T}}$

$$S_{km} = (\boldsymbol{V} \times \nabla)_{km} = -\epsilon_{ijm} V_{kj,i} = \begin{bmatrix} \frac{\partial V_{12}}{\partial x_3} - \frac{\partial V_{13}}{\partial x_2} & \frac{\partial V_{13}}{\partial x_1} - \frac{\partial V_{11}}{\partial x_3} & \frac{\partial V_{11}}{\partial x_2} - \frac{\partial V_{12}}{\partial x_1} \\ \frac{\partial V_{22}}{\partial x_3} - \frac{\partial V_{23}}{\partial x_2} & \frac{\partial V_{23}}{\partial x_1} - \frac{\partial V_{21}}{\partial x_3} & \frac{\partial V_{21}}{\partial x_2} - \frac{\partial V_{22}}{\partial x_1} \\ \frac{\partial V_{32}}{\partial x_3} - \frac{\partial V_{33}}{\partial x_2} & \frac{\partial V_{33}}{\partial x_1} - \frac{\partial V_{31}}{\partial x_3} & \frac{\partial V_{31}}{\partial x_2} - \frac{\partial V_{32}}{\partial x_1} \end{bmatrix} \quad \text{(B.7)}$$

Substituting the components $\boldsymbol{V}$ of as per Eq. (B.2),

$$S_{km} = \begin{bmatrix} \frac{\partial b_1}{\partial x_3} - \frac{\partial c_1}{\partial x_2} & \frac{\partial c_1}{\partial x_1} - \frac{\partial a_1}{\partial x_3} & \frac{\partial a_1}{\partial x_2} - \frac{\partial b_1}{\partial x_1} \\ \frac{\partial b_2}{\partial x_3} - \frac{\partial c_2}{\partial x_2} & \frac{\partial c_2}{\partial x_1} - \frac{\partial a_2}{\partial x_3} & \frac{\partial a_2}{\partial x_2} - \frac{\partial b_2}{\partial x_1} \\ \frac{\partial b_3}{\partial x_3} - \frac{\partial c_3}{\partial x_2} & \frac{\partial c_3}{\partial x_1} - \frac{\partial a_3}{\partial x_3} & \frac{\partial a_3}{\partial x_2} - \frac{\partial b_3}{\partial x_1} \end{bmatrix} \quad \text{(B.8)}$$

From Eq. (B.4), $-(Q_{km})^T$ is

$$-(Q_{km})^T = \begin{bmatrix} \frac{\partial b_1}{\partial x_3} - \frac{\partial c_1}{\partial x_2} & \frac{\partial c_1}{\partial x_1} - \frac{\partial a_1}{\partial x_3} & \frac{\partial a_1}{\partial x_2} - \frac{\partial b_1}{\partial x_1} \\ \frac{\partial b_2}{\partial x_3} - \frac{\partial c_2}{\partial x_2} & \frac{\partial c_2}{\partial x_1} - \frac{\partial a_2}{\partial x_3} & \frac{\partial a_2}{\partial x_2} - \frac{\partial b_2}{\partial x_1} \\ \frac{\partial b_3}{\partial x_3} - \frac{\partial c_3}{\partial x_2} & \frac{\partial c_3}{\partial x_1} - \frac{\partial a_3}{\partial x_3} & \frac{\partial a_3}{\partial x_2} - \frac{\partial b_3}{\partial x_1} \end{bmatrix} = S_{km} \tag{B.9}$$

# Appendix C

When Stokes' theorem is applied to higher order tensors, a transpose sign is introduced unlike in the case of a vector. The prove provided by Cermelli and Gurtin (Cermelli and Gurtin, 2001) is explained here in brief. For all constant vectors $\boldsymbol{c}$ and a tensor field $\boldsymbol{T}$, the identity $(\mathrm{curl}\,\boldsymbol{T}).\boldsymbol{c} = \mathrm{curl}(\boldsymbol{T}^T.\boldsymbol{c})$ holds. Now considering a smooth vector field $\boldsymbol{f}$, the Stokes' theorem applies as

$$\oint_C \boldsymbol{f}.\mathrm{d}\boldsymbol{x} = \iint_S \mathrm{curl}(\boldsymbol{f}).\boldsymbol{n}\mathrm{d}S \tag{C.1}$$

If $\boldsymbol{f} = \boldsymbol{T}^T.\boldsymbol{c}$, then

$$\oint_C \boldsymbol{T}^T.\boldsymbol{c}.\mathrm{d}\boldsymbol{x} = \iint_S \mathrm{curl}(\boldsymbol{T}^T.\boldsymbol{c}).\boldsymbol{n}\mathrm{d}S = \iint_S \mathrm{curl}(\boldsymbol{T}).\boldsymbol{c}.\boldsymbol{n}\mathrm{d}S \tag{C.2}$$

Since $\boldsymbol{c}$ is a constant therefore, taking transpose on both sides of Eq. (C.2)

$$\oint_C \boldsymbol{T}.\mathrm{d}\boldsymbol{x} = \iint_S (\mathrm{curl}\,\boldsymbol{T})^T.\boldsymbol{n}\mathrm{d}S \tag{C.3}$$

# Appendix D

## Micro-beam Laue diffraction

In micro-beam Laue diffraction, the recorded images correspond to the sum of the intensity scattered by the entire volume illuminated by the incident beam. Thus, the depth along the incident beam from which a specific diffraction signal originated is unknown. Hence, if several grains are illuminated simultaneously, or if there are large lattice distortions, the Laue spots become broadened and difficult to interpret. At the 34-ID-E instrument this limitation can be overcome by carrying out depth-resolved Laue measurements using the Differential Aperture X-ray Microscopy (DAXM) technique. Here, a ~50 μm diameter wire is scanned in small steps between the detector and the diffracting sample. The depth vs intensity profile for each pixel on the detector is calculated by subtracting the diffraction images from the consecutive wire position increments and triangulating using the wire edge and the line of the beam. A detailed description of the DAXM technique and the 34-ID-E instrument is provided elsewhere (Hofmann et al., 2013; Liu et al., 2010, 2004). Measurements were done to a depth of 20 μm beneath the sample surface. Laue diffraction patterns contained 30+ peaks and were indexed and fitted using the LaueGo software package (J.Z. Tischler: tischler@anl.gov) to extract both lattice orientation and the deviatoric elastic strain tensor at each measured point in 3D space. The measured strain and rotation gradients were then used to calculate the dislocation tensor and GND density.

# Appendix E

## 3D CPFE model

In the 3D CPFE model, the mechanical response of the tungsten BCC crystal under indentation was predicted using a constitutive law incorporating crystallographic slip. A brief description of the constitutive law, originally developed by Dunne et al. (Dunne et al., 2012, 2007), is provided here.

Recalling Eq. (22), it is known that the deformation gradient $\boldsymbol{F}$, splits multiplicatively into its elastic and plastic parts. Plastic deformation occurs on a slip system, $\lambda$, when the resolved shear stress $\tau^{\lambda}$ is greater than the critically resolved shear stress (CRSS). $\boldsymbol{F}^p$ can be defined in terms of the crystallographic slip $\boldsymbol{\beta}^p$ (relative displacement of two slip planes separated by a unit distance), slip direction $\boldsymbol{s}$ and slip plane normal $\boldsymbol{n}$. The crystallographic plane normals and the slip directions are updated as the crystal lattice undergoes deformation. For a finite number of slip systems, $\boldsymbol{F}^p$ is given by the sum of the contribution of the slip systems to the resultant slip

$$\boldsymbol{F}^P = \boldsymbol{I} + \frac{\partial \boldsymbol{u}^P}{\partial \boldsymbol{X}} = \boldsymbol{I} + \boldsymbol{\beta}^p = \boldsymbol{I} + \sum_\lambda {\beta^p}^\lambda \left(\boldsymbol{s}^\lambda \otimes \boldsymbol{n}^\lambda\right) \tag{E.1}$$

The rate of change of $\boldsymbol{F}^p$ is thus

$$\dot{\boldsymbol{F}^p} = \sum_\lambda \dot{{\beta^p}}^\lambda \left(\boldsymbol{s}^\lambda \otimes \boldsymbol{n}^\lambda\right) \tag{E.2}$$

where $\dot{{\beta^p}}^\lambda$ is the crystallographic slip rate on slip system $\lambda$. The velocity gradient $\boldsymbol{L}$ is given by

$$\boldsymbol{L} = \frac{\partial v}{\partial x} = \dot{\boldsymbol{F}}\boldsymbol{F}^{-1} \tag{E.3}$$

The velocity gradient can be split into symmetric and anti-symmetric components to give the rate of deformation $\boldsymbol{D}$ and the continuum spin $\boldsymbol{W}$ respectively. The total rate of deformation can be written as a sum of the elastic and plastic rates of deformation as

$$\boldsymbol{D} = \boldsymbol{D}^e + \mathrm{sym}\left(\boldsymbol{F}^e \boldsymbol{L}^p {\boldsymbol{F}^e}^{-1}\right) \cong \boldsymbol{D}^e + \mathrm{sym}(\boldsymbol{L}^p) \cong \boldsymbol{D}^e + \boldsymbol{D}^p \tag{E.4}$$

$\boldsymbol{D}^e$ is computed using Hooke's law, while $\boldsymbol{D}^p$ is approximated by the symmetric part of $\boldsymbol{L}^p$. $\boldsymbol{L}^p$ can be written as

$$\boldsymbol{L}^p = \dot{\boldsymbol{F}}^p \boldsymbol{F}^{p-1} = \sum_\lambda \dot{\beta^p}^\lambda (\boldsymbol{s}^\lambda \otimes \boldsymbol{n}^\lambda)\left(\boldsymbol{I} + \sum_\lambda {\beta^p}^\lambda (\boldsymbol{s}^\lambda \otimes \boldsymbol{n}^\lambda)\right)^{-1} \tag{E.5}$$

$$\cong \sum_\lambda \dot{\beta^p}^\lambda (\boldsymbol{s}^\lambda \otimes \boldsymbol{n}^\lambda)\left(\boldsymbol{I} - \sum_\lambda {\beta^p}^\lambda (\boldsymbol{s}^\lambda \otimes \boldsymbol{n}^\lambda)\right) \cong \sum_\lambda \dot{\beta^p}^\lambda (\boldsymbol{s}^\lambda \otimes \boldsymbol{n}^\lambda) \cong \dot{\boldsymbol{F}}^p$$

where, the higher products are ignored for small deformations.

When the UMAT is called by Abaqus, the UMAT is provided with the deformation gradient ($\boldsymbol{F}$) at the beginning and end of the time increment and the internal state variables at the beginning of the time increment. The UMAT returns the updated values of the state variables at the end of the time increment, the updated stress state ($\boldsymbol{\sigma}_{t+\Delta t}$) and the material Jacobian ($\frac{\partial \Delta \sigma}{\partial \Delta \varepsilon}$). $\boldsymbol{D}^p$ approximated from $\boldsymbol{L}^p$ (Eq. E4), gives the increment in the plastic strain ($\Delta \boldsymbol{\varepsilon}_p = \boldsymbol{D}^p \Delta t$), and the increment in the total strain can be computed from the known value the deformation gradient ($\Delta \boldsymbol{\varepsilon}$ = sym($\dot{\boldsymbol{F}}\, \boldsymbol{F}^{-1}$)$\Delta t$). $\boldsymbol{\sigma}_{t+\Delta t}$ can then be written in terms of a trial stress ($\boldsymbol{\sigma}_{tr}$) and a plastic corrector term as follows

$$\boldsymbol{\sigma}_{t+\Delta t} = \boldsymbol{C}' (\boldsymbol{\varepsilon}_t^e + \Delta \boldsymbol{\varepsilon}^e) = \boldsymbol{C}' (\boldsymbol{\varepsilon}_t^e + \Delta \boldsymbol{\varepsilon} - \Delta \boldsymbol{\varepsilon}^p) = \boldsymbol{\sigma}_{tr} - \boldsymbol{C}' \Delta \boldsymbol{\varepsilon}^p \tag{E.6}$$

Where $\boldsymbol{C}'$ is the stiffness matrix rotated into the sample coordinate system. The UMAT used here solves these equations implicitly, i.e. all quantities are written at the end of the time increment and the stress is forced to converge back onto the yield surface within a tolerance of $10^{-12}$ MPa using the plastic corrector term $\boldsymbol{C}' \Delta \boldsymbol{\varepsilon}^p$. The reduction of this stress residual ($\Psi = \boldsymbol{\sigma}_{t+\Delta t} - \boldsymbol{\sigma}_{tr} + \boldsymbol{C}' \Delta \boldsymbol{\varepsilon}^p$) is done using the Newton-Raphson iterative method. Further details can be found in (Dunne et al., 2007).

The physically based slip law used here determines the slip rate on the slip system by considering the thermally activated process of movement of dislocations, overcoming pinning obstacles. For a slip system with average dislocation glide speed $\langle v \rangle^\lambda$ Burgers' vector magnitude $b^\lambda$, with $q$ dislocations per unit area h-L as shown in Figure E.1(a), the crystallographic slip rate maybe written using the Orowan equation

$$\dot{\beta}^{p^{\lambda}} = q\frac{\langle v\rangle^{\lambda} b^{\lambda}}{Lh} = \rho_m^{\lambda} b^{\lambda} \langle v\rangle^{\lambda} \tag{E.7}$$

where, $\rho_m^{\lambda}$ is the density of mobile dislocations.

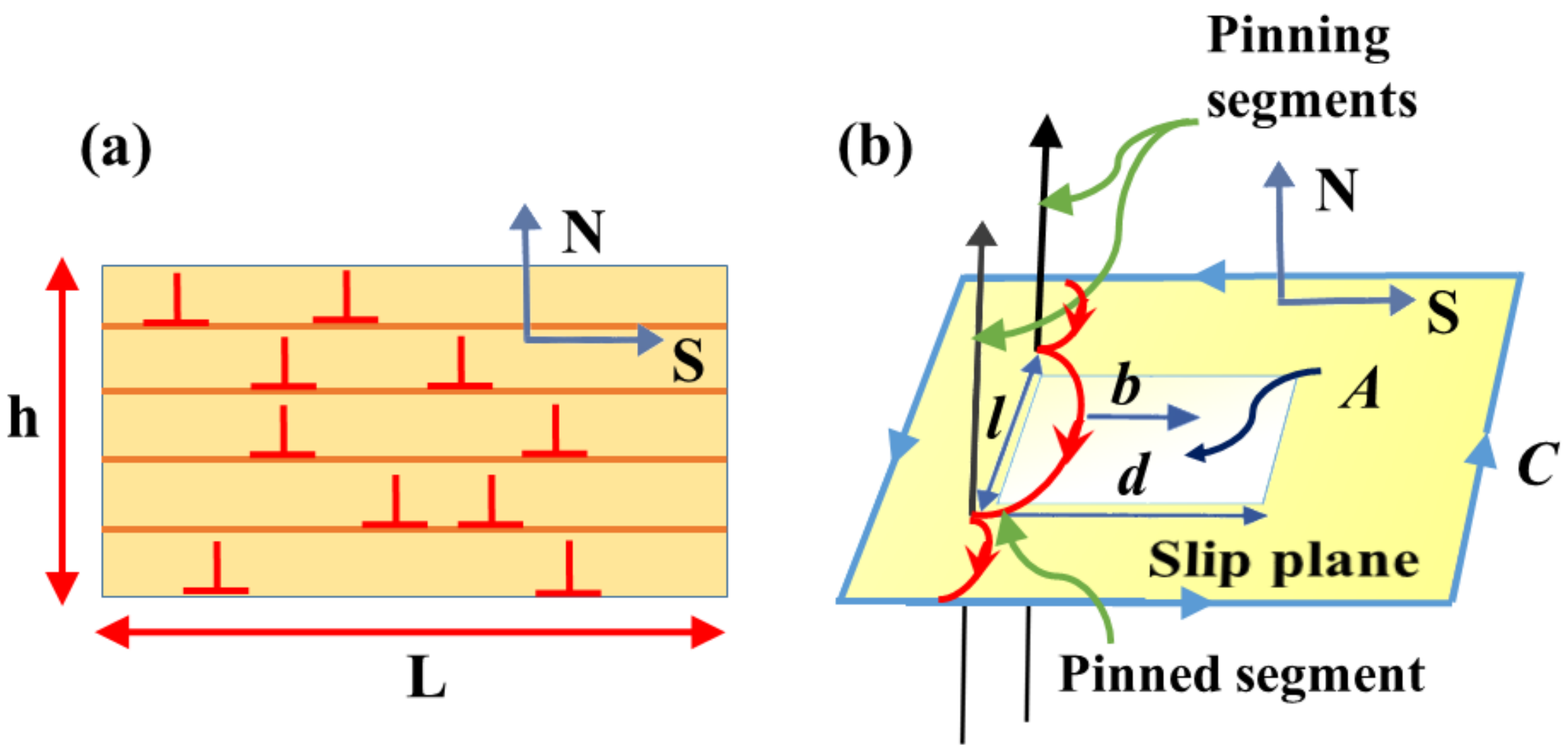

Figure E.1 - (a) Schematic diagram of a set of slip planes, viewed edge on, each comprising of a random distribution of dislocations on one slip system (b) One of the slip planes from (a), viewed in cross-section, with slip plane normal ***N*** and slip direction ***S***. *C* represents the closed circuit path around the slip plane used to determine the Burgers' vector of the black cutting dislocations. An edge dislocation (shown in red), pinned by cutting dislocations from another slip system, where *l* is the distance between the pinning points and *d* the distance jumped by the dislocation on overcoming the pinning dislocations.

The thermal activation process influences the glide velocity by enabling the pinned dislocations to overcome the energy barriers produced by obstacles. Figure E.1 (b) shows an edge dislocation, pinned by cutting dislocations from another slip system, where $l$ is the distance between the pinning points and $d$ the distance jumped by the dislocation on overcoming the pinning dislocations. The average glide velocity is given by

$$\langle v \rangle = d\Gamma \tag{E.8}$$

where, $\Gamma$ is the rate of escape of dislocations given by

$$\Gamma = \frac{\nu b^{\lambda}}{2l} \exp\left(-\frac{\Delta G}{kT}\right) \tag{E.9}$$

where, $v$ is the frequency of dislocation jumps, $G$ is the Gibbs free energy, $k$ the Boltzmann constant and $T$ the temperature in Kelvin. In terms of Helmholtz free energy ($\Delta F$) and applied stress field $\boldsymbol{\tau}$, the Gibbs free energy is

$$\Delta G = \Delta F - \tau V \tag{E.10}$$

where, $V$ is the activation volume. Substituting this expression of $\Delta G$ into Eq. (E.8) and taking into account the forward and backward activation, and the critically resolved shear stress $\tau_c$, the slip rate for a slip system $\lambda$, maybe written as,

$$\dot{\beta^p}^{\lambda} = \rho_g \nu \left(b^{\lambda}\right)^2 \exp\left(-\frac{\Delta F^{\lambda}}{kT}\right) \sinh\left(\frac{\mathrm{sgn}(\tau^{\lambda})\left(\left|\tau^{\lambda}\right| - \tau_c^{\lambda}\right)V^{\lambda}}{kT}\right) \tag{E.11}$$

$V$ depends on the spacing between the pinning dislocations $l$. A lengthscale dependent mechanical response arises as the strain rate and subsequent plastic deformation reduces as the GND density increases. We assume for simplicity,

$$l = \frac{1}{\sqrt{\Psi(\rho_{GND} + \rho_{SSD})}} \tag{E.12}$$

where, $\Psi$ is a coefficient representing the probability of pinning.

The values of the properties in the constitutive law have been acquired from literature and are listed in Table E.1.

| Material Property | Value | Reference |
|---|---|---|
| Elastic modulus $E$ | 421 GPa | (Ayres et al., 1975; Bolef and De Klerk, 1962; Featherston and Neighbours, 1963; Klein and Cardinale, 1993) |
| Shear modulus $G$ | 164.4 GPa | (Ayres et al., 1975; Bolef and De Klerk, 1962; Featherston and Neighbours, 1963; Klein and Cardinale, 1993) |
| Poisson's ratio $\upsilon$ | 0.28 | (Ayres et al., 1975; Bolef and De Klerk, 1962; Featherston and Neighbours, 1963; Klein and Cardinale, 1993) |
| Burgers' vector $b$ | $2.7 \times 10^{-10}$ m | (Dutta and Dayal, 1963) |
| Helmholtz free energy $\Delta H$ | $3.4559 \times 10^{-20}$ J | (Kartal et al., 2015) |
| Boltzmann constant $k$ | $1.381 \times 10^{-23}$ J/K | (Sweeney et al., 2013) |
| Temperature $T$ | 293 K | Room temperature assumed similar to experimental conditions |
| Attempt frequency $v$ | $1 \times 10^{-11}$ s$^{-1}$ | (Sweeney et al., 2013) |
| Density of statistically stored dislocations, $\rho_{SSD}$ | $1 \times 10^{10}$ m$^{-2}$ | (Sweeney et al., 2013) |
| CRSS, $\tau_c$ | 900 MPa | Fitted to data |
| Density of mobile dislocations $\rho_m$ | $5 \times 10^{10}$ m$^{-2}$ | (Kartal et al., 2015) |
| Probability of pinning $\Psi$ | $1.457 \times 10^{-4}$ | (Kartal et al., 2015) |

Table E.1 – Material properties for tungsten taken from literature and the CRSS from fitting to load-displacement experimental data obtained from nano-indentation.

# Appendix F

## Expected lattice rotations

As the block deforms due to indentation, it is expected that lattice rotations about the X-axis, will be positive and negative (right-handed rotation) on either side of the indent centre (depicted by the dotted blue line). Similarly, it can be expected that $\theta_y$ will be negative in the region labelled before indent centre and positive after the indent centre. Rotations about the Z-axis are anticipated to be small.

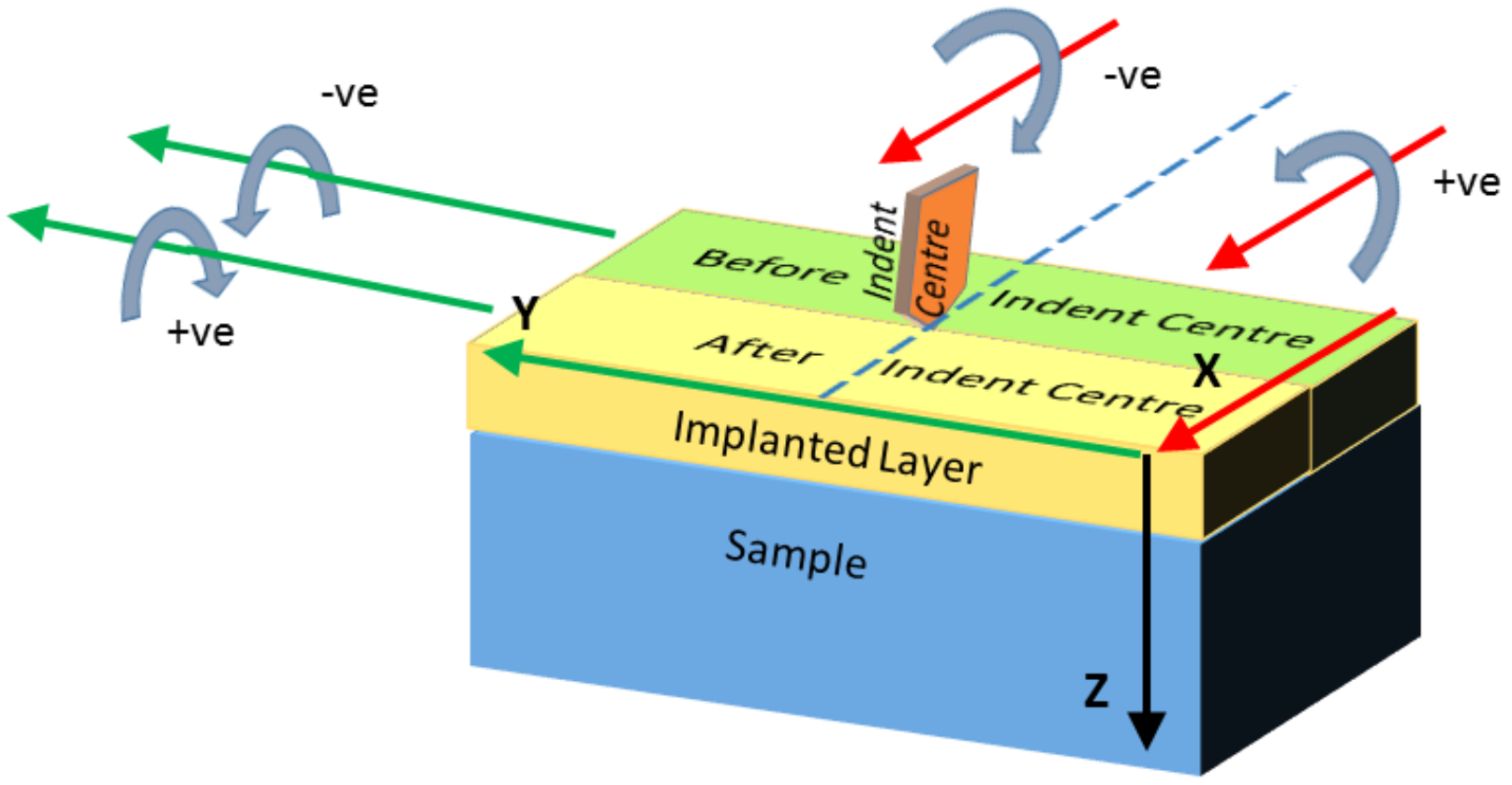

Figure F.1–Schematic of the lattice rotations expected due to nano-indentation

# Appendix G

Table G.1 shows the combination of Burgers' vector and line direction used for GND density calculation.

| Dislocation type | Density | Burgers' vector : $\boldsymbol{b}$ | Slip Normal : $\boldsymbol{n}$ | Line Direction : $\boldsymbol{l}$ |
|---|---|---|---|---|
| Edge 1 | $\rho_1$ | [-111] | (110) | [-1 1 -2] |
| Edge 2 | $\rho_2$ | [1-11] | (110) | [-1 1 2] |
| Edge 3 | $\rho_3$ | [111] | (-101) | [1 -2 1] |
| Edge 4 | $\rho_4$ | [1-11] | (-101) | [-1 -2 -1] |
| Edge 5 | $\rho_5$ | [111] | (-110) | [-1 -1 2] |
| Edge 6 | $\rho_6$ | [11-1] | (-110) | [1 1 2] |
| Edge 7 | $\rho_7$ | [111] | (01-1) | [-2 1 1] |
| Edge 8 | $\rho_8$ | [-111] | (01-1) | [-2 -1 -1] |
| Edge 9 | $\rho_9$ | [1-11] | (011) | [-2 -1 1] |
| Edge 10 | $\rho_{10}$ | [11-1] | (011) | [2 -1 1] |
| Edge 11 | $\rho_{11}$ | [-111] | (101) | [1 2 -1] |
| Edge 12 | $\rho_{12}$ | [11-1] | (101) | [1 -2 -1] |
| Screw 1 | $\rho_{13}$ | [-111] | (110) | [-1 1 1] |
| Screw 2 | $\rho_{14}$ | [1-11] | (110) | [1 -1 1] |
| Screw 3 | $\rho_{15}$ | [111] | (-101) | [1 1 1] |
| Screw 4 | $\rho_{16}$ | [11-1] | (-110) | [1 1 -1] |

Table G.1 - Combination of Burgers' vector and line direction used for the calculation of the GND

**Acknowledgements**

We thank Dr. E. Elmukashfi for insightful discussions about the curl computation, Dr. L. Hansen for providing the spherical indenter tip, Prof. M. Rieth for providing single crystal tungsten samples, and Prof. F. Dunne for providing a copy of his UEL code. This research used resources of the Advanced Photon Source, a U.S. Department of Energy (DOE) Office of Science User Facility operated for the DOE Office of Science by Argonne National Laboratory under Contract No. DE-AC02-06CH11357.

**Funding**

This work was funded by Leverhulme Trust Research Project Grant RPG-2016-190. ET acknowledges the Engineering and Physical Sciences Research Council, fellowship grant EP/N007239/1.